\documentclass[11pt]{article}
\usepackage{amsmath,amssymb,amsfonts} 
\usepackage{graphics}                 
\usepackage{makeidx}
\usepackage{graphicx}
\usepackage{sidecap}

\usepackage[caption=false,font=normalsize,labelfont=sf,textfont=sf]{subfig}
\usepackage{algorithmic}
\usepackage[algoruled,vlined]{algorithm2e}
\usepackage{fullpage}
\usepackage{float}

\numberwithin{equation}{section}
\numberwithin{figure}{section}
\numberwithin{table}{section}

\newtheorem{theorem}{Theorem}[section]
\newtheorem{proposition}[theorem]{Proposition}
\newtheorem{corollary}[theorem]{Corollary}

\newtheorem{remark}[theorem]{Remark}
\newtheorem{definition}[theorem]{Definition}
\def \srr{\stackrel{\mathrm{def}}{=}}
\newcommand{\rf}[1]{(\ref{#1})}

\def \eh{Eq. \rf}

\def \a{\alpha}
\def \n{\nu}
\def \k{\kappa}
\def \b{\beta}
\def \w{\omega}
\def \la{\lambda}
\def \br{\begin{remark}}
\def \er{\end{remark}}
\def \bdd{\begin{definition} }
\def \bpp{\begin{proposition} }
\def \bcc{\begin{corollary} }

\def \edd{\end{definition}}
\def \epp{\end{proposition}}
\def \ecc{\end{corollary} }

\begin{document}
\title{Image inpainting using directional  wavelet packets originating from  polynomial  splines }

\author{Amir Averbuch$^1$~~Pekka Neittaanm\"aki$^2$~~Valery  Zheludev$^1$~~Moshe Salhov$^1$ ~~Jonathan Hauser$^3$  \\
$^1$School of Computer Science, $^3$School of Electrical Engineering\\
Tel Aviv University, Israel\\
$^2$Faculty of Mathematical Information Technology\\
 University of Jyv\"askyl\"a, Finland}

 \date{ }
\maketitle
\begin{abstract}
The paper presents  a new algorithm for the image inpainting  problem. The algorithm  uses  a recently designed versatile library of   quasi-analytic   complex-valued wavelet packets (qWPs) which originate from polynomial  splines of arbitrary orders.  Tensor products of 1D qWPs  provide a diversity of 2D qWPs oriented in multiple directions. For example, a set of the fourth-level qWPs comprises 62 different directions. The properties of these  qWPs such as refined frequency resolution,  directionality of waveforms with unlimited number of orientations,  (anti-)symmetry of waveforms and windowed oscillating structure of waveforms with a variety of frequencies, make them  efficient in image processing applications, in particular, in dealing with  the inpainting problem addressed in the paper. The obtained results for this problem are quite competitive with the best state-of-the-art algorithms. The inpainting is implemented by an iterative scheme, which expands  the \emph{Split Bregman Iteration} (SBI) procedure by supplying it with an  adaptive variable soft thresholding  based on the  \emph{Bivariate Shrinkage} algorithm. In the inpainting experiments,  performance comparison between the  qWP-based  methods and the state-of-the-art algorithms is presented.
\end{abstract}

\section{Introduction}\label{sec:s1}
Multimedia images as well as biomedical, seismic and hyper-spectral images, to name a few, comprise smooth regions, oriented  edges in various directions and texture that can have an oscillating structure. One of the main goals of  image processing  is  to recover the image from a degraded input data. The degradation can result, for example,  from missing  pixels, noise, blurring  and from different combinations of these factors.  Another goal is the extraction of a limited number of characteristic features from images for  pattern recognition and machine learning tasks.  Achievement of the above goals relies on the fact that practically all processed images  have a sparse representation in a proper transform  domain. A sparse representation of an image means that the image can be approximated by a linear combination of a relatively small number of 2D ``basic" elements from some dictionary, while retaining the above mentioned components of the image. The dictionary of such elements should comprise waveforms that
\begin{enumerate}
  \item Are oriented in multiple directions (for edges capturing);
  \item Have an oscillating structure with multiple frequencies (for retaining texture patterns);
  \item Have vanishing moments, at least locally (for sparse representation of smooth regions);
  \item Good localization   in the spatial domain;
  \item Produce a refined frequency  separation.
  \item   Fast implementation of the corresponding transforms is important.
\end{enumerate}

In  recent years, a number of elements'  dictionaries, which meet some of the above requirements, were described such as   pseudo-polar processing \cite{averbuch2008frameworkI,averbuch2008frameworkII},  contourlets \cite{Contour}, curvelets \cite{curve,curve1} and  shearlets \cite{kuty,shear}. These dictionaries were used in various image processing  applications. For example,  one    successful application of
 shearlet transforms to  the image processing  is the methodology based on the  \emph{Digital Affine Shear transforms} (DAS-1) \cite{zhuang}.  However, while successfully capturing edges in images, those  dictionaries did not demonstrate a satisfactory  texture restoration due to  the lack of oscillatory waveforms in the dictionaries' libraries.

Another approach to design directional  dictionaries consists of the tensor multiplication of complex wavelets (\cite{king1,barakin}), wavelet  frames and wavelet packets (WPs) (\cite{jalob1,bay_sele,bhan_zhao,bhan_com_sup, bhan_zhao_zhu}, to name a few).  The tight tensor-product complex  wavelet  frames (TP\_$\mathbb{C}$TF$_{n}$) with different number of  directions, are designed in \cite{bhan_zhao,bhan_zhao_zhu,bhan_com_sup} (here \emph{n} refers to the number of filters
in the underlying 1D complex tight framelet filter bank). Some of them, in particular TP\_$\mathbb{C}$TF$_{6}$ and its modifications  like cptTP\_$\mathbb{C}$TF$_{6}$ with compactly supported framelets and TP\_$\mathbb{C}$TF$^{\downarrow}_{6}$ with the reduced redundancy, demonstrate impressive performance for image denoising and inpainting. The waveforms in these frames are oriented in 14 directions and, due to the 2-layer structure of their spectra possess some oscillatory properties.

Some of the disadvantages in the above 2D  TP\_$\mathbb{C}$TF$_{6}$, cptTP\_$\mathbb{C}$TF$_{6}$  and TP\_$\mathbb{C}$TF$^{\downarrow}_{6}$ frames such as, for example, limited and fixed number of  directions (14  directions at each decomposition  level) are overcome in
\cite{che_zhuang} using the algorithm   \emph{Digital Affine Shear Filter Transform with 2-Layer Structure (DAS-2)}. It is done  by the incorporation of the two-layer structure, which is inherent in the TP\_$\mathbb{C}$TF$_{6}$  frames, into directional filter banks introduced in \cite{zhuang}. This  incorporation improves the  performance of DAS-2 compared to  TP\_$\mathbb{C}$TF$_{6}$ on texture-rich images such as ``Barbara", which is not the case on smoother images like ``Lena".

Recently, we  designed a family of dictionaries  that maximally meet the requirements 1--6  \footnote{The design of the  dictionaries and implementation of the corresponding transforms are described in detail in \cite{azn_pswq}. } and utilize them in image processing  applications. Building blocks for such a design are  orthonormal  WPs originated from   discretized  polynomial  splines  of multiple orders (see Chapter 4 in \cite{ANZ_book3}).
The designed  dictionaries consist of 2D quasi-analytic   directional wavelet packets (qWPs). The design scheme is briefly outlined in Section \ref{sec:s2}.

The qWPs  are applied to image inpainting to restore images from corrupted data where many pixels are missing and additive noise is present.  This paper introduces two qWP-based iterative methods:  Method1 (\textbf{M1}) and Method2 (\textbf{M2}) to deal with the inpainting problem. The performances of \textbf{M1} and \textbf{M2} are compared  with the performance of   \textbf{Algorithm I} introduced in \cite{braver1}  (see  Section \ref{sec:s3} )  that uses  the SET-4\footnote{ SET-4 is the set of   filter banks  DAS-2,  DAS-1, TP-$\mathbb{C}$TF$_6$ and  TP-$\mathbb{C}$TF$_6^{\downarrow}$.}
filter banks.   The computational scheme in \textbf{M1} is close to the scheme in \textbf{Algorithm I}.  The difference is that  the qWP transforms are used instead of the SET-4 filter banks. \textbf{M2} couples \textbf{M1} with  the \emph{split
Bregman iteration} (SBI) scheme, which uses the
so-called \emph{\aa-based} approach (\cite{gold_os,ji_shen_xu}).
Practically, in all the conducted experiments,
  \textbf{M1} and especially  \textbf{M2}  outperform the DET-4 algorithms in the  Structural Similarity Index (SSIM) sense. In many experiments, especially on texture rich images, \textbf{M2} produces higher  PSNR values than all the SET-4 algorithms.

The paper is organized as follows.
In order for the paper to be self-contained, Section \ref{sec:s2} briefly outlines the design and implementation of the  qWP transforms in one and two dimensions.  Section \ref{sec:s3} describes the qWP-based \textbf{M1} and \textbf{M2}  for image inpainting. Section \ref{sec:s4}  presents the experimental results from the restoration of  images degraded by missing up to 80\% of its pixels while additive noise is present. The performances of the qWP-based methods are compared with the performances of the algorithms that use the SET-4 filter banks.  Section \ref{sec:s5} discusses the results.

Notations and abbreviations are given in Table \ref{nota}

\begin{table}
\resizebox{\textwidth}{!}{
  \centering
  \begin{tabular}{|l|l|l|}
     \hline
     $N=2^{j}$,\quad $\delta[n]$-- $N$-periodic  Kronecker delta& $\w\srr e^{2\pi\,i/N}$ &  $\cdot^{\ast}$ -- complex conjugate\\
     $\Pi[N]$ -- space of   $N$-periodic  signals &\textbf{M1} -- Method1 &\textbf{M2} -- Method2  \\
$\Pi[N,N]$-- space of 2D   $N$-periodic   arrays&2D --  two-dimensional& WP -- wavelet packet\\
     DFT --  Discrete Fourier  transform  &FFT -- Fast Fourier  transform    & WPT -- WP transform \\
cWP   -- complimentary WP & qWP   -- quasi-analytic    WP &HT -- Hilbert transform \\
     $H(\mathbf{x})$ -- HT of $\mathbf{x}\in\Pi[N]$ & DTS --  discrete-time spline & DTSWP -- WP from DTS \\
     BSA -- Bivariate Shrinkage algorithm    &p-filter  -- periodic  filter  & SBI  -- split Bregman iteration \\
    SSIM -- Structural Similarity Index  & SET-4 -- filter banks:  DAS-2,     & DAS-1, TP-$\mathbb{C}$TF$_6$,  TP-$\mathbb{C}$TF$_6^{\downarrow}$\\
    PSNR -- Peak Signal-to-Noise ratio& in decibels (dB) &$10\log_{10}\left(\frac{N\,255^2}{\sum_{k=1}^N(x_{k}-\tilde
  x_{k})^2}\right)\; dB$.\\
    \hline
   \end{tabular}
   }
  \caption{Notations and abbreviations}\label{nota}
\end{table}

\section{Quasi-analytic   WPs}\label{sec:s2}
In this section, we  briefly outline the design and properties of   quasi-analytic    WPs (qWPs). For details see \cite{azn_pswq}.
 Notations for this section are listed in Table \ref{nota21}.

\begin{table}
\resizebox{\textwidth}{!}{
  \centering
  \begin{tabular}{|l|l|}
     \hline
     $ u^{p}[n]$ and  $ v^{p}[n]$ -- DFTs of sampled B-spline --\eh {psp_char}& sequences $\a[n], \;\b[n]$ -- Eq.\rf{on_dsDFT}\\
     ${\psi}_{[m],l}^{p}$ -- DTSWP of order $p$ No. $l$ from level $m$ -- Eqs.\rf{on_dsDFT}, \rf{mlev_wq})& $\tilde{\mathbf{H}}={\mathbf{H}}=\left\{{\mathbf{h}}_{0},{\mathbf{h}}_{1}\right\}$ -- p-filter banks\\
$\mathbf{H}_{[m]}=\left\{\mathbf{h}_{[m],0},\mathbf{h}_{[m],1}\right\}$ --$m$-level p-filter bank , \eh{irfg_m}&  $\psi_{[m],j,l}^{p}[k,n]\srr \psi_{[m],j }^{p}[k]\, \psi_{[m],l}^{p}[n]$-- 2D DTSWPs \\
 \hline
   \end{tabular}
   }
  \caption{Notations for Section \ref{sec:ss21}}\label{nota21}
\end{table}
\subsection{Orthonormal WPs originated from polynomial  splines}\label{sec:ss21}

In this section, we list the main properties of  periodic  discrete-time  wavelet packets originated from polynomial  splines  (DTSWPs) and the  corresponding transforms. For details see Chapter 4 in \cite{ANZ_book3}.
\subsubsection{One-dimentional DTSWPs}\label{sec:sss211}
The centered  $N$-periodic   polynomial  B-spline $B^{p}(t)$ of order $p$ is an $N$-periodization of the function
$ {b}^{p}(t)=\frac{1}{(p-1)!}\sum_{k=0}^{p}(-1)^{k}\,{p\choose k}\,\left(t+\frac{p}{2}-k\right)_{+}^{p-1}$ where $x_{+}\srr\max\{x,0\}.$
The B-spline $B^{p}(t)$ is supported on the interval $(-p/2,p/2)$ up to periodization. It is strictly positive inside this interval and  symmetric about zero, where it has its single maximum and has $p-2$ continuous derivatives.
The functions $
  {S}^{p}(t) \srr \sum_{l=0}^{N-1}q[l]\,{B}^{p}[t-l]
$
 are referred to as  the order-$p$  periodic  splines. The following two sequences will be used later:
\begin{eqnarray}\label{psp_char}
   u^{p}[n]\srr\sum_{k=-N/2}^{N/2-1}\w^{-kn}\,{b}^{p}\left(k\right),\quad
v^{p}[n]\srr\w^{-n/2}\sum_{k=-N/2}^{N/2-1}\w^{-kn}\,{b}^{p}\left(k+\frac{1}{2}\right).
\end{eqnarray}

 The two-times dilation of the B-spline  $b^{p}(t)$ is denoted by $b_{d}^{p}(t)=b^{p}(t/2)/2$.
\bdd\label{bds_defT}The span-two discrete-time  B-spline $\mathbf{b}_{[1]}^{p}$ of order $p$  is defined as an $N$-periodization of the sampled B-spline $b_{d}^{p}(t)$. It is  an $N$-periodic  signal  from $\Pi[N]$:
\begin{equation*}\label{bts2_uv}
 {b}^{p}_{[1]}[k]\srr b_{d}^{p}(k),\;k=-N/2,...,N/2-1(\mathrm{mod}\, N),\quad\hat{b}^{p}_{[1]}[n]= \frac{u^{p}[2n]+v^{p}[2n]}{2}.
\end{equation*}
\edd

Linear combinations   ${s}_{[1]}^{p}[k]=\sum_{l=0}^{N/2-1}q[l]\,{b}_{[1]}^{p}[k-2l]$ of two-sample shifts of the B-splines are referred to as periodic  discrete-time splines (DTSs) of span 2.  Their DFTs are  $\hat{{s}}_{[1]}^{p}[n]=\hat{q}[n]_{1}\,  \hat{{b}}^{p}_{[1]}[n] $.
The $N/2$-dimensional  space of the DTSs  is denoted by $^{p}{{\mathcal{S}}}_{[1]}^{0}\subset \Pi[N]$ and its orthogonal  complement in the signal   space   $\Pi[N]$   is denoted by $^{p}{{\mathcal{S}}}_{[1]}^{1}$ .

Define the  DTS  ${\psi}_{[1],0}^{p}$ and the signal  ${\psi}_{[1],1}^{p}$ by their DFTs:
\begin{equation}\label{on_dsDFT}
      \hat{ \psi}_{[1],0}^{p}[n]=\b[n]\srr 2\frac{u^{p}[2n]+  v^{p}[2n]}{\sqrt{u^{p}[2n]^{2}+  v^{p}[2n]^{2}}},\quad  \hat{{ \psi}}_{[1],1}^{p}[n]= \a[n]\srr2\w^{n}\,\frac{u^{p}[2n]-  v^{p}[2n]}{\sqrt{u^{p}[2n]^{2}+  v^{p}[2n]^{2}}}.
     \end{equation}
    Two-sample shifts of the discrete-time-spline    ${\psi}_{[1],0}^{p}[k]$ and of  the signal  ${\psi}_{[1],1}^{p}[k]$  form  orthonormal  bases of the subspaces  $\,{}^{p}{{\mathcal{S}}}_{[1]}^{0}$ and $\,{}^{p}{{\mathcal{S}}}_{[1]}^{1}$, respectively.

     The  signals $ {\psi}_{[1],0}^{p}[k]$ and $ {\psi}_{[1],1}^{p}[k]$   are real-valued and
     symmetric about $k_{0}=0$ and $k_{1}=-1$, respectively.  They are referred to as the discrete-time-spline    wavelet packets (DTSWPs) of order $p$ from the first decomposition  level

The one-level DTSWP transform  of a signal  $\mathbf{x}\in\Pi[N]$ and its inverse are implemented by the application of the analysis  $\tilde{\mathbf{H}}=\left\{\tilde{\mathbf{h}}_{0},\tilde{\mathbf{h}}_{1}\right\}$ and  synthesis  ${\mathbf{H}}=\left\{{\mathbf{h}}_{0},{\mathbf{h}}_{1}\right\}$ p-filter banks to the signal  $\mathbf{x}$ and the transform  coefficients, respectively. The synthesis  p-filters are ${\mathbf{h}}_{s}=\tilde{\mathbf{h}}_{s},\;s=0,1.$ Their impulse responses and frequency responses are
\begin{equation}\label{irfg}
  {h}_{s}[k]={ \psi}_{[1],s}^{p}[k],\qquad \hat{h}_{0}[n]=\b[n],\quad\hat{h}_{1}[n]=\a[n].
\end{equation}
 The transforms can be
represented in a matrix form:
   \begin{equation*}\label{mod_repAS1}
    \left(
     \begin{array}{c}
       \hat{y}_{[1]}^{0}[n] \\
         \hat{y}_{[1]}^{1}[n]\\
     \end{array}
   \right)=\frac{1}{2}\tilde{\mathbf{M}}[-n]\cdot \left(
     \begin{array}{l}
      \hat{x}[n] \\
       \hat{x}[\vec{n}]
     \end{array}
   \right),\quad  \left(
     \begin{array}{l}
      \hat{x}[n] \\
       \hat{x}[\vec{n}]
     \end{array}
   \right)={\mathbf{M}}[n]\cdot \left(
     \begin{array}{c}
        \hat{y}_{[1]}^{0}[n] \\
         \hat{y}_{[1]}^{1}[n]\\
     \end{array}
   \right),
    \end{equation*}
    where $\vec{n}=n+{N}/{2}$ and $\tilde{\mathbf{M}}[n]$ and ${\mathbf{M}}[n]$ are the  modulation matrices of the analysis  and synthesis  p-filter banks:
  \begin{eqnarray}\label{sa_modma10T}
    {\mathbf{{M}}}[n]=\sqrt{2}\left(
                                 \begin{array}{cc}
                                   {\b}[n] & {\a}[n]   \\
                             {\b}\left[n+\frac{N}{2}\right]     & {\a}\left[n+\frac{N}{2}\right]  \\
                                 \end{array}
                               \right)= \tilde{\mathbf{{M}}}[n]^{T}.
  \end{eqnarray}
The sequences $\b[n]$ and $\a[n]$ are defined in \eh{on_dsDFT}.

\paragraph{Extension of transforms to deeper decomposition  levels.}
\par\noindent
The p-filter bank  is defined as $\mathbf{H}_{[m]}=\left\{\mathbf{h}_{[m],0},\mathbf{h}_{[m],1}\right\}$, where the p-filters $\mathbf{h}_{[m],s}, \;s=0,1,\; m=1,...,M,$ are defined by their frequency responses:
\begin{equation}\label{irfg_m}
 \hat{h}_{[m],0}[n]=\hat{h}_{0}[2^{m}n]=\b[2^{m}n],\quad \hat{h}_{[m],1}[n]=\hat{h}_{1}[2^{m}n]=\a[2^{m}n].
\end{equation}
The WP transforms to deeper decomposition  levels are implemented iteratively,  while the transform  coefficients $\left\{\mathbf{y}_{[m+1]}^{r}\right\}$
 are derived by filtering the coefficients $\left\{\mathbf{y}_{[m]}^{l}\right\}$ with the time-reversed p-filters $\mathbf{h}_{[m],s},$ where $l=0,...,2^m-1,\;s=0,1$ and $r=\left\{
                                                                                                                               \begin{array}{ll}
                                                                                                                                 2l +s, & \hbox{if $l$ is even;} \\
                                                                                                                                 2l +(1-s), & \hbox{if $l$ is odd.}
                                                                                                                               \end{array}
                                                                                                                             \right.
$

The transform  coefficients are ${y}_{[m]}^{l}[lk]=\left\langle \mathbf{x},\psi^{p}_{[m],l}[\cdot, -2^{m}k] \right\rangle$, where the signals $\psi^{p}_{[m],l}$ are normalized, orthogonal  to each other in the space $\Pi[N]$, and their $2^{m}k-$sample shifts are mutually orthogonal. They are referred to as the level-$m$ DTSWPs of order $p$. The set $\left\{\psi^{p}_{[m],l}[\cdot, -2^{m}k] \right\},\;l=0,...,2^m-1,\;k=0,...N/2^m-1,$ constitutes an orthonormal  basis of the  space $\Pi[N]$ and generates its split into $2^m$ orthogonal  subspaces. The next-level WPs $\psi^{p}_{[m+1],r}$ are derived by filtering the wavelet packets  $\psi^{p}_{[m],l}$ with the p-filters $\mathbf{h}^{s}_{[m]}$ such that
\begin{equation}\label{mlev_wq}
  {\psi}_{[m+1],r}^{p}[n]  =\sum_{k=0}^{N/2^{m}-1}{h}_{[m],s}[k] \, {\psi}_{[m],l}^{p}[n-2^{m}k].
\end{equation}
The   DTSWPs  originated from ninth-order splines and their spectra are displayed in Fig. \ref{psi_phiT}.
The transforms are executed in the spectral domain  with the  Fast Fourier transform  (FFT) by  using the modulation matrices $\tilde{\mathbf{M}}[2^{m}\,n]$ and ${\mathbf{M}}[2^{m}\,n]$ of the p-filter bank  $\mathbf{H}_{[m]}$.

\subsubsection{Two-dimentional DTSWPs}\label{sec:sss212}
A standard way to extend  one-dimensional (1D) WPTs to multiple dimensions is by tensor-product extension.
The 2D one-level WPT of a  signal  $\mathbf{x}=\left\{x[k,n]\right\},\;k,n=0,...,N-1,$  which  belongs to $\Pi[N,N]$, consists of the application of  1D WPT to columns of  $\mathbf{x}$, which is followed by the application of the transform  to  rows of the  coefficients array. As a result of the 2D WPT of signals from $\Pi[N,N]$, the space  split
 into four mutually orthogonal  subspaces
$ \Pi[N,N]=\bigoplus_{j,l=0}^{1}\,^{p} {\mathcal{S}}^{j,l}_{[1]}.$

The 2D DTSWPs are $\psi_{[1],j,l}^{p}[k,n]\srr \psi_{[1],j }^{p}[k]\, \psi_{[1],l}^{p}[n],   \quad j,l=0,1.$ They are normalized and  orthogonal  to each other in the space $\Pi[N,N]$. It means that  $$\sum_{k,n=0}^{N-1}\psi_{[1],j_{1} ,l_{1}}^{p}[k,n]\,\psi_{[1],j_{2} ,l_{2}}^{p}[k,n]=\delta[j_{1}-j_{2}]\,\delta[l_{1}-l_{2}].$$
 Their two-sample shifts in both directions are mutually orthogonal.
The subspace ${}^{p} {\mathcal{S}}^{j,l}_{[1]}$ is a linear hull of two-sample shifts of the 2D wavelet packets
$\left\{\psi_{[1],j,l}^{p}[\cdot-2k,\cdot-2n]\right\} ,\;k,n,=0,...,N/2-1,$ that form an orthonormal  basis of ${}^{p} {\mathcal{S}}^{j,l}_{[1]}$.

By the application of the above transforms iteratively to blocks of the transform  coefficients down to $m$-th level, we get that the space $ \Pi[N,N]$ is decomposed into $4^{m}$ mutually orthogonal  subspaces
$ \Pi[N,N]=\bigoplus_{j,l=0}^{2^{m}-1}\,^{p} {\mathcal{S}}^{j,l}_{[m]}.$

The 2D tensor-product  wavelet packets $\psi_{[m],j ,l}^{p}\srr \psi_{[m],j }^{p}[k]\, \psi_{[m],l}^{p}[n]$, which provide orthonormal  bases to the respective subspaces $^{p} {\mathcal{S}}^{j,l}_{[m]},$ are well localized in the spatial domain, their 2D DFT spectra  produce a refined tiling of the frequency  domain of signals from $ \Pi[N,N].$\footnote{ It is true especially for WPs derived from  higher-order DTSs.} The drawback for image processing   is that the WPs are  oriented  in ether horizontal or vertical directions or are not oriented  at all.

\subsection{Quasi-analytic and complementary WPs}\label{sec:ss22}
In this section, we define the so-called quasi-analytic    WPs related to the DTSWPs discussed in Section \ref{sec:ss21}. In addition,  we introduce an orthonormal  set of waveforms which are complementary to the DTSWPs.
Notations for this section are listed in Table \ref{nota22}.

\begin{table}
\resizebox{\textwidth}{!}{
  \centering
  \begin{tabular}{|l|l|}
     \hline
      $\tau^{p}_{[m],l}=H(\psi^{p}_{[m],l})$ -- HT  of DTSWP $\psi^{p}_{[m],l}$  &${\psi}^{p}_{\pm[m],l}=\psi^{p}_{[m],l}   \pm i\tau^{p}_{[m],l}$ -- analytic   DTSWPs \\
$\hat{\phi}^{p}_{[m],l} $  -- complimentary WP (cWP) --\eh{phi_df} & $\Psi^{p}_{\pm[m],l}$   -- complex quasi-analytic    WPs (qWPs) --\eh{qaz}  \\
$\mathbf{Z}_{[m]}$ -- $m$-level transform  coefficients with the qWPs $\left\{\Psi_{\pm[m]}\right\}$, \eh{coeLD1}& $\mathbf{H},\, \mathbf{F},\, \mathbf{Q}_{\pm}$ -- p-filter banks for qWP transforms, \eh{hfq} \\
    $\Psi_{++[m],j , l}^{p}[k,n] $ and  $\Psi_{+-[m],j , l}^{p}[k,n] $ -- 2D complex qWPs, \eh{coq2} &$\theta_{+[m],j ,l}^{p}[k,n]$ and $\theta_{-[m],j ,l}^{p}[k,n]$ -- directional  qWPs, \eh{vt_pm}   \\
    \hline
   \end{tabular}
   }
  \caption{Notations for Section \ref{sec:ss22}}\label{nota22}
\end{table}

Briefly, the design scheme  consists of the following steps: 1. Application of the Hilbert transform  (HT) to the set $\left\{\psi\right\}$ of  orthonormal  WPs  producing the set  $\left\{\tau=H(\psi)\right\}$.
2. A slight correction of the  set  $\left\{\tau\right\}$ provides an orthonormal  set   $\left\{\phi\right\}$ of the so-called complimentary WPs (cWPs), which are anti-symmetric and whose magnitude spectra coincides with the magnitude spectra of the respective WPs from the  set $\left\{\psi\right\}$. 3. Two sets of complex quasi-analytic   WPs (qWPs) $\left\{\Psi_{+}\srr\psi+i\,\phi\right\}$ and $\left\{\Psi_{-}\srr\psi-i\,\phi\right\}$, whose spectra are localized in the positive and negative half-bands of the frequency  domain, respectively, are defined. 4. Two sets of 2D complex qWPs produced by the tensor products of the qWPs $\left\{\Psi_{\pm}\right\}$ as: $\left\{\Psi_{++}\srr\Psi_{+}\bigotimes\Psi_{+}\right\}$ and $\left\{\Psi_{+-}\srr\Psi_{+}\bigotimes\Psi_{-}\right\}$ are defined. 5. The dictionaries we use are  the real parts of these qWPs: $\left\{\theta_{\pm}\srr\mathfrak{Re}(\Psi_{+\pm})\right\}$.

\begin{SCfigure}
\centering
\caption{Diagram of the qWP design (left) and quadrants of  frequency  domain (right)}
\includegraphics[width=2.7in]{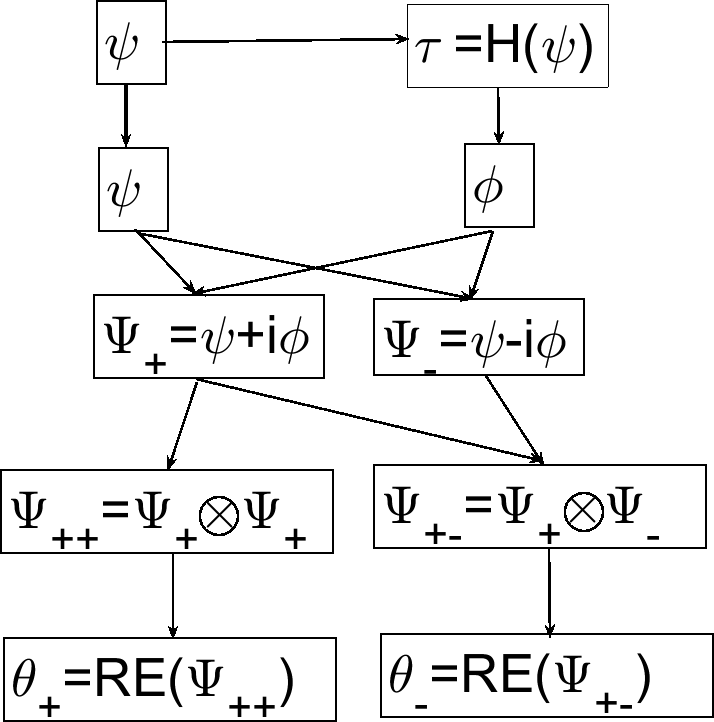}\quad \vline\quad  \includegraphics[width=1.5in]{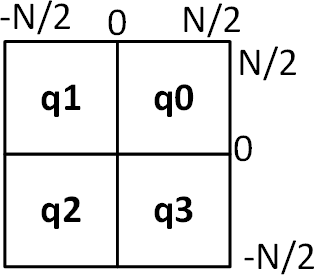}
  \label{dia_fipsi}
\end{SCfigure}

The diagram in Fig. \ref{dia_fipsi} illustrates the design of  qWPs.

\subsubsection{ Analytic periodic  signals}\label{sec:sss221}

A signal  $\mathbf{x}\in\Pi[N]$ is represented by its  inverse DFT which can be written as follows:
\begin{eqnarray*}\label{fsx+}
  x[k]&=&\frac{\hat{x}[0]+(-1)^{k}\hat{x}[N/2]}{N}+\frac{2}{N}\sum_{n=1}^{N/2-1}\frac{\hat{x}[n]\,\w^{kn}+(\hat{x}[n]\,\w^{kn})^{\ast}}{2}.
\end{eqnarray*}
Define the real-valued  signal  $\mathbf{t}\in\Pi[N]$ and two complex-valued signals $\mathbf{{x}}_{+}$ and  $\mathbf{{x}}_{-}$ such that
\begin{equation*}
\label{yy}
\begin{array}{lll}
 t[k]&\srr&\frac{2}{N}\sum_{n=1}^{N/2-1}\frac{\hat{x}[n]\,\w^{kn}-\hat{x}[n]^{\ast}\,\w^{-kn}}{2i},\\
 {x}_{\pm}[k]&\srr&x[k]\pm i\,t[k]=\frac{\hat{x}[0]+(-1)^{k}\hat{x}[N/2]}{N}\\&+&\frac{2}{N}\sum_{n=1}^{N/2-1}
 \left\{
   \begin{array}{ll}
     \hat{x}[ n]\,\w^{ kn}, & \hbox{for $\bar{x}_{+}$;} \\
      \hat{x}[- n]\,\w^{- kn}=\hat{x}[ N-n]\,\w^{- k(N-n)}, & \hbox{for $\bar{x}_{-}$.}
   \end{array}
 \right.
\end{array}
\end{equation*}

The spectrum of $\mathbf{{x}}_{+}$ comprises only non-negative frequencies and vice versa for $\mathbf{{x}}_{-}$\footnote{Recall that $\hat{x}[N/2]=\hat{x}[-N/2]$.}. We have
$\mathbf{x}=\mathfrak{Re}(\mathbf{{x}}_{\pm})$ and $\mathfrak{Im}(\mathbf{{x}}\pm)=\pm\mathbf{t}$. The  signals  $\mathbf{{x}}_{\pm}$ are referred to as the  periodic  discrete-time  analytic signals.

The signal  $\mathbf{t}$   is a discrete  periodic  version of the  Hilbert transform  (HT) of a discrete-time  periodic  signal   $\mathbf{x}$, that is  $\mathbf{t}=H(\mathbf{x})$ (see \cite{opp}, for example). Note that the DFT spectrum $\hat{t}[n]$ lacks the values $\hat{t}[0]$ and $\hat{t}[N/2]$.

\subsubsection{Analytic  and quasi-analytic   WPs}\label{sec:sss222}
The analytic   WPs and their DFT spectra are  derived from  DTSWPs $\left\{\psi^{p}_{[m],l}\right\},\;m=1,...,M,\;l=0,...,2^{m}-1,$ in line with the scheme in Section \ref{sec:sss221}. Recall that for all $l\neq0$, the DFT $\hat{\psi}^{p}_{[m],l}[0]=0$ and  for all $l\neq2^{m}-1$, the DFT $\hat{\psi}^{p}_{[m],l}[N/2]=0$.

Denote by $\tau^{p}_{[m],l}\srr H(\psi^{p}_{[m],l})$ the HT  of the wavelet packet  $\psi^{p}_{[m],l}$.
Then, the  analytic   WPs are
\(
{\psi}^{p}_{\pm[m],l}=\psi^{p}_{[m],l}   \pm i\tau^{p}_{[m],l}.
\)

\paragraph{Complementary orthonormal   WPs}
The values $\hat{\tau}^{p}_{[m],l}[0]$ and $\hat{\tau}^{p}_{[m],l}[N/2]$ are missing in the DFT spectra of the HT signals $\tau^{p}_{[m],l}$.
We upgrade the set $\left\{\tau^{p}_{[m],l}\right\},\;l=0,...2^{m}-1,$ in the following way. The  set $\left\{\phi^{p}_{[m],l}\right\},\;m=1,...,M, \;l=0,...,2^{m}-1,$  of signals from the space $\Pi[N]$ is  defined
via their DFTs:
\begin{equation}\label{phi_df}
  \hat{\phi}^{p}_{[m],l}[n]= \hat{\tau}^{p}_{[m],l}[n]+
                 \hat{\psi}^{p}_{[m],l}[0]+\hat{\psi}^{p}_{[m],l}[N/2].
\end{equation}
For all $l\neq0, 2^{m}-1,$ the signals $\phi^{p}_{[m],l}$ coincide with $\tau^{p}_{[m],l}=H(\psi^{p}_{[m],l})$.
\bpp\label{pro:phi_oo}
\par\noindent
\begin{description}
  \item[-] The magnitude spectra  $\left|\hat{\phi}^{p}_{[m],l}[n]\right|$  of cWPs coincide with $\left|\hat{\psi}^{p}_{[m],l}[n]\right|$.
  \item[-] For any   $m=1,...,M,$ and $l=1,...,2^{m}-2,$ the signals \ $\phi^{p}_{[m],l}$ are antisymmetric oscillating waveforms.   For $l= 0, \,2^{m}-1$, the shapes of the signals are near antisymmetric.
      \item[-] The orthonormality properties that are similar to the properties of WPs  $\psi^{p}_{[m],l}$ hold for the signals  $\phi^{p}_{[m],l}$ such that
      $        \left\langle\phi^{p}_{[m],l}[\cdot -k\,2^{m}],\phi^{p}_{[m],\la}[\cdot -n\,2^{m}] \right\rangle= \delta[\la,l]\,\delta[k,n],\quad k,n=0,...,N/2^{m}-1 .
     $
\end{description}
\epp

\begin{figure}
\centering
\includegraphics[width=5.5in]{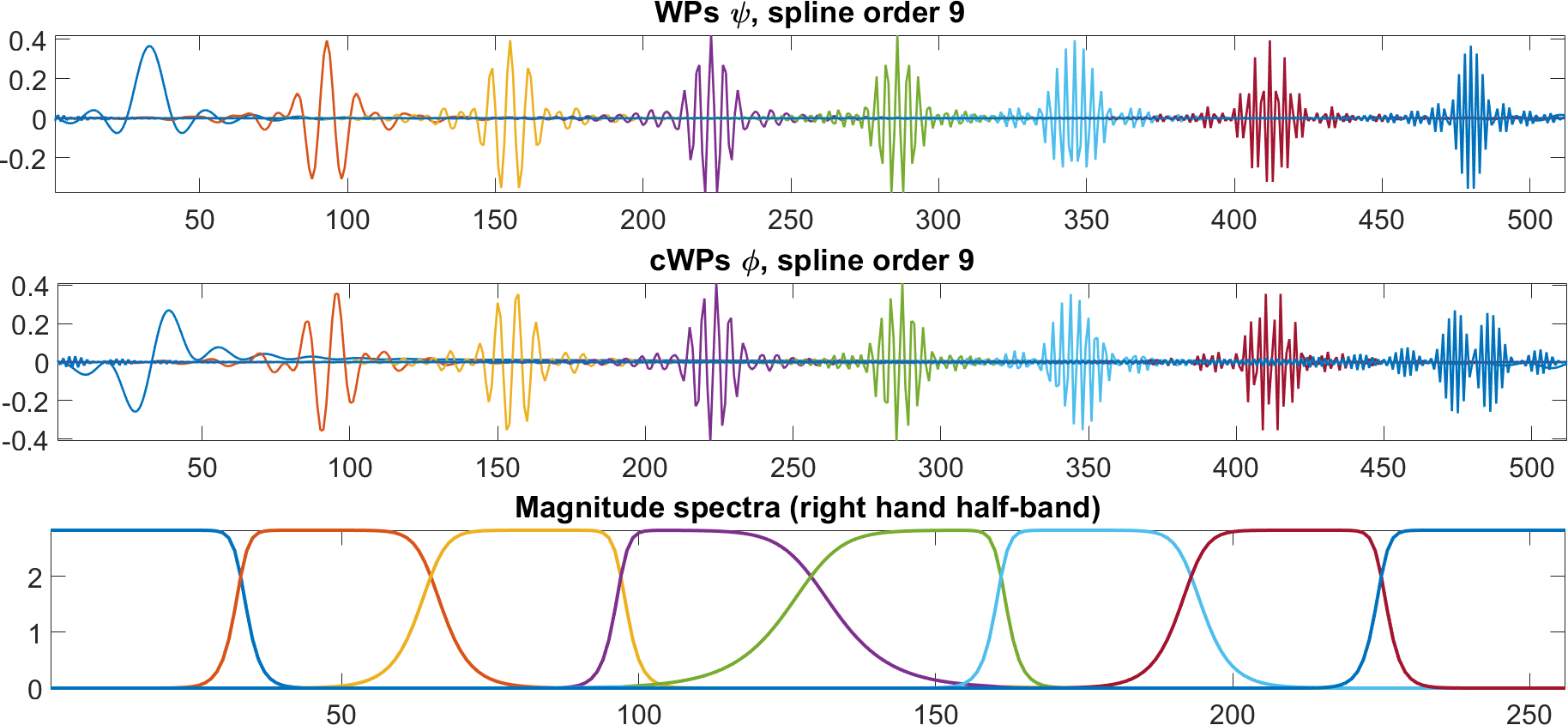}
\caption{Top: signals ${\psi}^{9}_{[3],l},\;l=0,...,7$. Center: signals ${\phi}^{9}_{[3],l},\;l=0,...,7$.  Bottom:  their magnitude DFT spectra, respectively}
    \label{psi_phiT}
\end{figure}

  Figure \ref{psi_phiT} displays  the signals ${\psi}^{9}_{[3],l}$ and ${\phi}^{9}_{[3],l},\;l=0,...,7$, from the third decomposition  level and their magnitude spectra.  Addition of $\hat{\psi}^{p}_{[3],l}[0]$ and $\hat{\psi}^{p}_{[3],l}[N/2]$ to the  spectra of ${\phi}^{9}_{[3],l},\;l=0,7$ results in an antisymmetry distortion.

We call the signals     $\left\{\phi^{p}_{[m],l}\right\},\;m=1,...,M, \;l=0,...,2^{m}-1$, the \emph{complementary  wavelet packets} (cWPs). Similarly to the WPs $\left\{\psi^{p}_{[m],l}\right\},$ differenent combinations of the cWPs can provide differenent  orthonormal  bases for the space $\Pi[N]$.
\paragraph{Quasi-analytic    WPs:}
The sets of complex-valued WPs, which we refer to as the quasi-analytic    wavelet packets (qWP),  are defined by
\begin{equation}\label{qaz}
  \Psi^{p}_{\pm[m],l}=\psi^{p}_{[m],l}   \pm i\phi^{p}_{[m],l}, \quad m=1,...,M,\;l=0,...,2^{m}-1,
\end{equation}
where $\phi^{p}_{[m],l}$ are the cWPs from \eh{phi_df}.  qWPs $\Psi^{p}_{\pm[m],l}$ differ from the analytic   WPs ${\psi}^{p}_{\pm[m],l}$ by the addition of the two values $\pm i\,\hat{\psi}^{p}_{[m],l}[0]$ and $\pm i\,\hat{\psi}^{p}_{[m],l}[N/2]$ into their DFT spectra, respectively. For a given decomposition  level $m$, these values are zero for all $l$ except for $l_{0}=0$ and $l_{m}=2^{m}-1$. It means that for all $l$ except for $l_{0}$ and $l_{m}$, the qWPs $\Psi^{p}_{\pm[m],l}$ are analytic.
The DFTs of qWPs are
\begin{eqnarray*}\label{qa_df}
  \hat{\Psi}^{p}_{+[m],l}[n]=\left\{
               \begin{array}{ll}
  (1+i)\hat{\psi}^{p}_{[m],l}[n], & \hbox{if $n=0, N/2$;} \\
                  2\hat{\psi}^{p}_{[m],l}[n], & \hbox{if $0< n< N/2$;} \\
               0 & \hbox{if $ N/2<n<N$,}
               \end{array}
             \right.\quad
 \hat{\Psi}^{p}_{-[m],l}[n]=\left\{
               \begin{array}{ll}
(1-i)\hat{\psi}^{p}_{[m],l}[n], & \hbox{if $n=0, N/2$;} \\
                  0 & \hbox{if $0< n< N/2$;} \\
               2\hat{\psi}^{p}_{[m],l}[n], & \hbox{if $ N/2< n< N$.}
               \end{array}
             \right.
\end{eqnarray*}
\subsubsection{Design of 1D cWPs and qWPs}\label{sec:sss223}
The DFTs of the first-level DTSWPs are
\(
  \hat{\psi}^{p}_{[1],0}[n]=\b[n],\quad  \hat{\psi}^{p}_{[1],1}[n]=\w^{n}\,\b[n+N/2]=\a[n],
\)
where the sequences  $\b[n] $ and  $\a[n] $ are defined in \eh{on_dsDFT}. The DFTs of  the  first-level cWPs are
\begin{eqnarray*}\label{df_cwq1}
  \hat{\phi}^{p}_{[1],0}[n]=\left\{
               \begin{array}{ll}
                  -i\,\b[n], & \hbox{if $0<n<N/2$;} \\
                i\, \b[n], & \hbox{if $N/2<n<N$;} \\
                 \sqrt{2}, & \hbox{if $n=0$; }
             \\
                 0, & \hbox{ if $n=N/2,$}
               \end{array}
\right.\quad \hat{\phi}^{p}_{[1],1}[n]=\left\{
               \begin{array}{ll}
                  -i\,\a[n], & \hbox{if $0<n<N/2$;} \\
                i\, \a[n], & \hbox{if $N/2<n<N$;} \\
                 0, & \hbox{if $n=0$; }
             \\
                 -\sqrt{2}, & \hbox{ if $n=N/2.$}
               \end{array}
\right.
\end{eqnarray*}
The cWPs and qWPs from the second and further decomposition  levels are derived iteratively using the same p-filter banks as the DTFWPs.

\bpp[\cite{azn_pswq}]\label{pro:cwq_des} Assume that for a DTSWP $\psi_{[m+1],r}^{p}$ the relation in \eh{mlev_wq} holds. Then, for the cWP $\phi_{[m+1],r}^{p}$ and  qWP $\Psi_{\pm[m+1],r}^{p}$ we have
\begin{eqnarray*}\label{mlev_cwq}
{\phi}_{[m+1],r}^{p}[k]  &=&\sum_{j=0}^{N/2^{m}-1}{h}_{[m],s}[j] \, {\phi}_{[m],l}^{p}[k-2^{m}j]\Longleftrightarrow\hat{\phi}_{[m+1],r}^{p}[n]= \hat{h}_{s}[2^{m}n]_{m}\,\hat{\phi}_{[m],l}^{p}[n],\\
 \Psi_{\pm[m+1],r}^{p}[k] & =&\sum_{j=0}^{N/2^{m}-1}{h}_{[m],s}[j] \, {\Psi}_{\pm[m],l}^{p}[k-2^{m}j]\Longleftrightarrow\hat{\Psi}_{\pm[m+1],r}^{p}[n]= \hat{h}_{s}[2^{m}n]_{m}\,\hat{\Psi}_{[m],l}^{p}[n],\\
\hat{h}_{[1]}^{0}[\n] &=&\hat{\psi}^{p}_{[1],0}[\n]=\b[\n],\quad  \hat{h}_{[1]}^{1}[\n] = \hat{\psi}^{p}_{[1],1}[\n]=\a[\n],\quad r=\left\{
                                                                                                                               \begin{array}{ll}
                                                                                                                                 2l +s, & \hbox{if $l$ is even;} \\
                                                                                                                                 2l +(1-s), & \hbox{if $l$ is odd.}
                                                                                                                               \end{array}
                                                                                                                             \right.
\end{eqnarray*}
\epp
\subsubsection{Design of  2D directional WPs}\label{sec:sss224}
The 2D wavelet packets,  whicht are defined by the tensor products of 1D DTSWPs such that
\(
  \psi_{[m],j ,l}^{p}[k,n]=\psi_{[m],j}^{p}[k]\,\psi_{[m], l}^{p}[n]
\) and cWPs \(
  \phi_{[m],j ,l}^{p}[k,n]=\phi_{[m],j}^{p}[k]\,\phi_{[m], l}^{p}[n] \), possess many valuable properties but they    lack  the directionality,   which is needed in many  applications that process 2D data.  However, real-valued 2D WPs oriented in multiple directions  are
derived from  tensor  products of complex  qWPs $\Psi_{\pm[m],l}^{p}$. The complex 2D qWPs are defined  as follows:
\begin{equation}\label{coq2}
\Psi_{++[m],j , l}^{p}[k,n] \srr \Psi_{+[m],j}^{p}[k]\,\Psi_{+[m], l}^{p}[n], \quad
  \Psi_{+-[m],j ,l}^{p}[k,n] \srr \Psi_{+[m],j}^{p}[k]\,\Psi_{-[m], l}^{p}[n],
\end{equation}
where $  m=1,...,M,\;j ,l=0,...,2^{m}-1,$ and $k ,n=-N/2,...,N/2-1$.
The real  parts of these 2D qWPs are
\begin{equation}
\label{vt_pm}
\begin{array}{lll}
 \theta_{+[m],j ,l}^{p}[k,n] &\srr& \mathfrak{Re}(\Psi_{++[m],j ,l}^{p}[k,n]) =  \psi_{[m],j ,l}^{p}[k,n]-\phi_{[m],j ,l}^{p}[k,n], \\
\theta_{-[m],j ,l}^{p}[k,n] &\srr&  \mathfrak{Re}(\Psi_{+-[m],j ,l}^{p}[k,n]) =  \psi_{[m],j ,l}^{p}[k,n]+\phi_{[m],j ,l}^{p}[k,n],\\
\end{array}
\end{equation}

\begin{figure}
\centering
\includegraphics[width=3.1in]{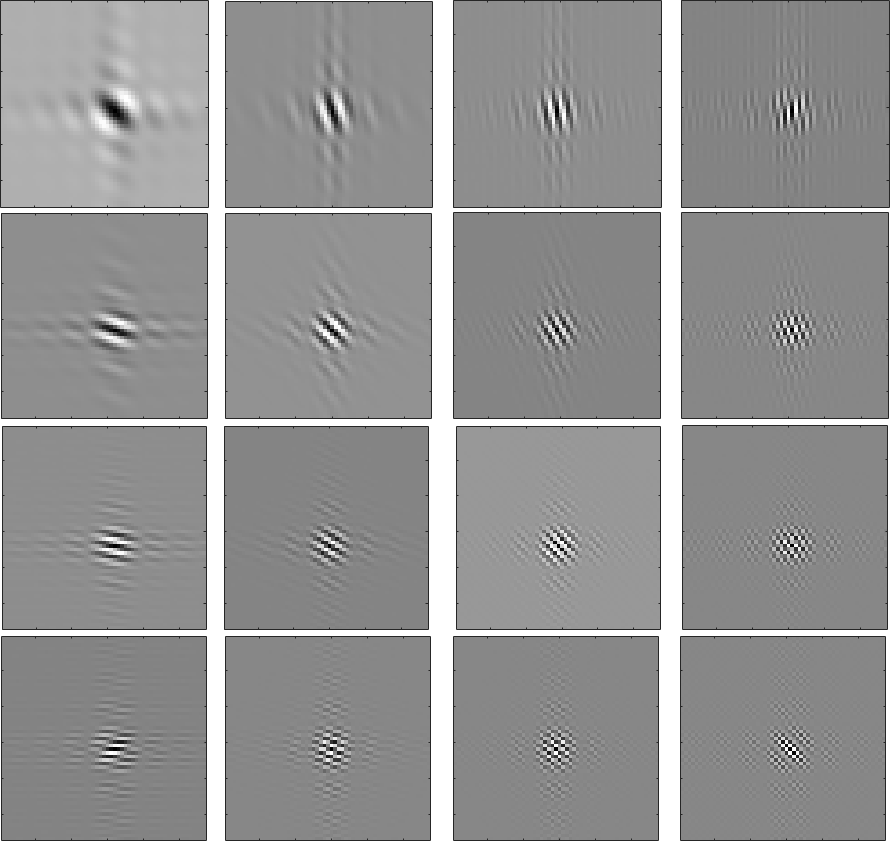}
\quad
\includegraphics[width=3.05in]{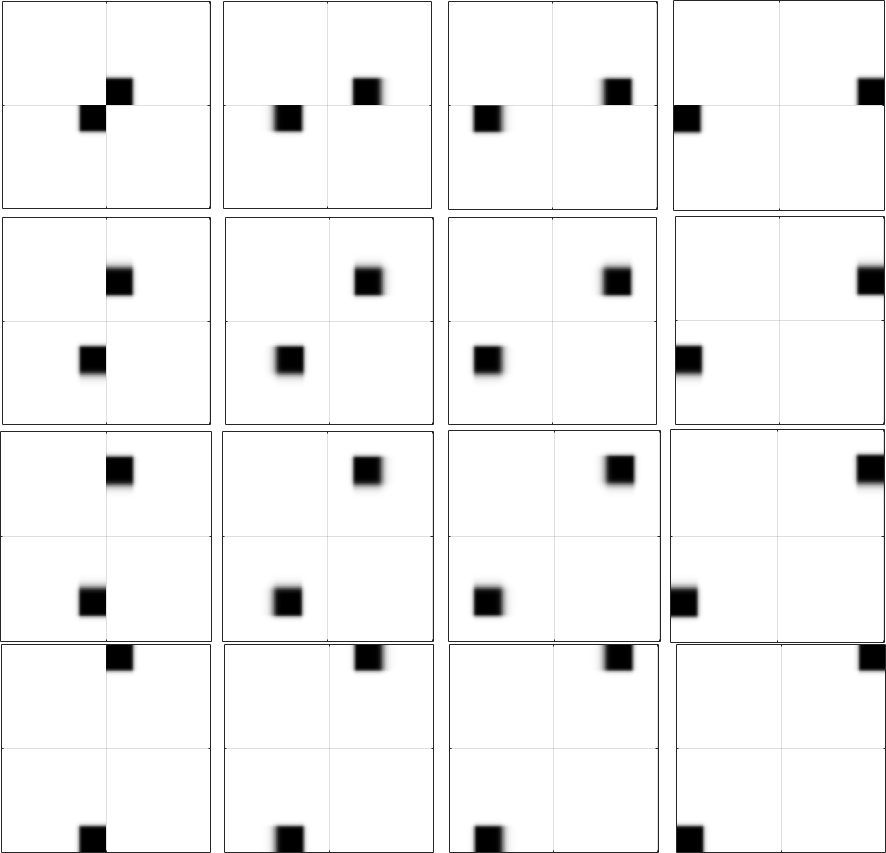}
\caption{WPs $\theta_{+[2],j ,l}^{9}$ from the second decomposition  level and their magnitude spectra}
\label{pp_2_2d}
\end{figure}

\begin{figure}
\centering
\includegraphics[width=3.1in]{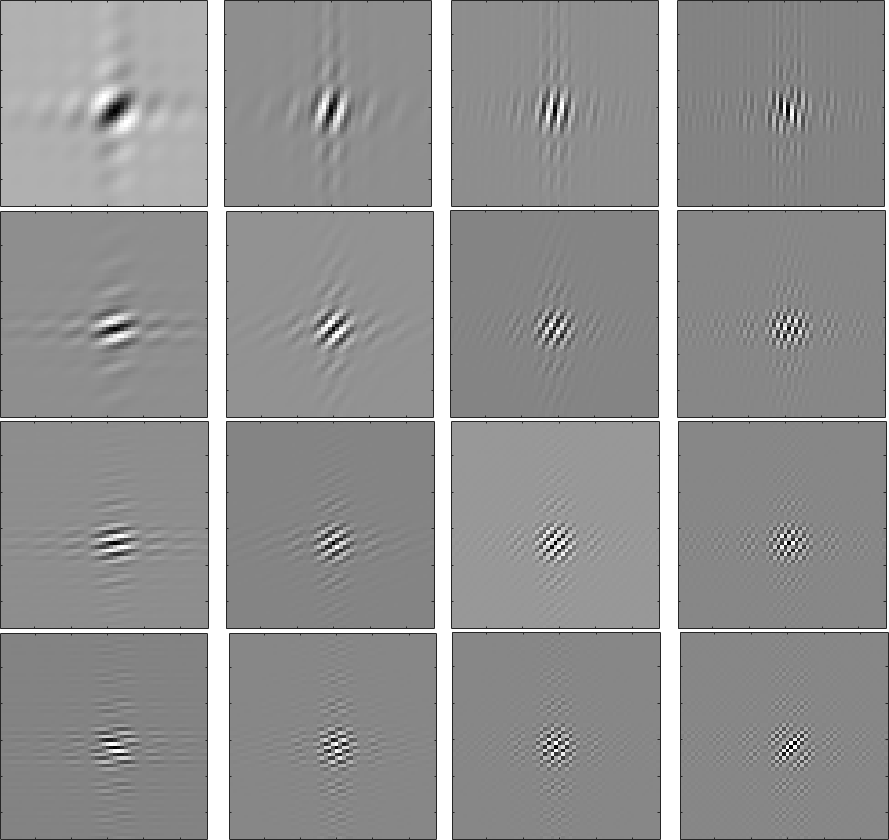}
\quad
\includegraphics[width=3.1in]{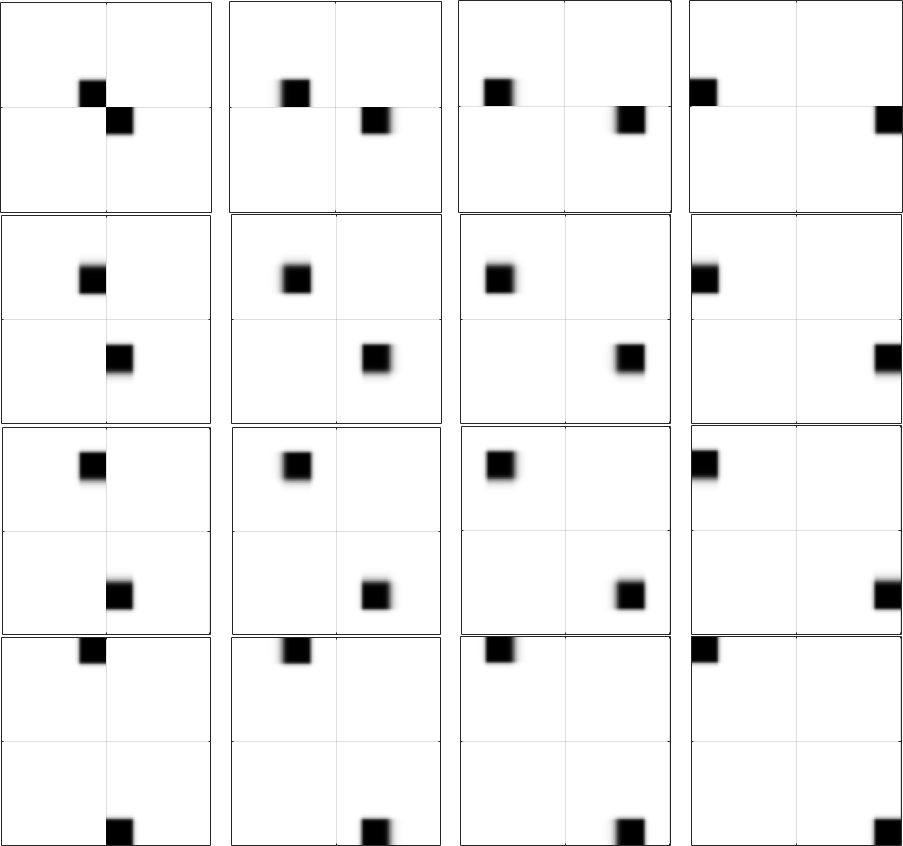}
\caption{WPs $\theta_{-[2],j ,l}^{9}$ from the second decomposition  level and their magnitude spectra}
\label{pm_2_2d}
\end{figure}

The DFT spectra of the 2D qWPs $\Psi_{++[m],j ,l}^{p},\;j ,l=0,...,2^{m}-1,$ are the  tensor products of the one-sided spectra of the qWPs:
\(
\hat{ \Psi}_{++[m],j ,l}^{p}[\k,\n] =\hat{ \Psi}_{+[m],j}^{p}[\k]\,\hat{\Psi}_{+[m], l}^{p}[\n]
\)
and, as such,  they tile the  quadrant $\mathbf{q}_{0}$ of the frequency  domain, while the spectra of $\Psi_{+-[m],j ,l}^{p},\;j ,l=0,...,2^{m}-1,$ tile the  quadrant $\mathbf{q}_{1}$ (see  Fig. \ref{dia_fipsi}). Consequently, the spectra of the real-valued 2D qWPs $ \theta_{+[m],j ,l}^{p}$ and $ \theta_{-[m],j ,l}^{p}$  tile the  pairs of  quadrants $\mathbf{q}_{+}\srr\mathbf{q}_{0}\bigcup\mathbf{q}_{2}$ and $\mathbf{q}_{-}\srr\mathbf{q}_{1}\bigcup\mathbf{q}_{3}$, respectively, by relatively small pairs of squares, which are symmetric about the origin.  Such localizations and shapes  of the spectra  determine the directionality and shapes of the waveforms $ \theta_{\pm[m],j ,l}^{p}$. They are, approximately, windowed cosines with multiple frequencies, which are  oriented, at the level $m$, in  $2(2^{m+1}-1)$ directions (\cite{azn_pswq}).
Figures \ref{pp_2_2d} and  \ref{pm_2_2d} display real qWPs  $\theta_{+[2],j ,l}^{9}$  and $\theta_{-[2],j ,l}^{9},\;j,l=0,1,2,3,$ respectively, from the second decomposition  level and their magnitude spectra.
\\ \\

\subsubsection{Implementation of  qWP transforms}\label{sec:sss225}
The qWP transforms are executed in the frequency  domain using the FFT. Assume that the signal  \textbf{x } to be transformed belongs to the space $\Pi[N]$ of $N-$periodic  discrete-time  signals.

The sets $\mathbf{Z}_{[m]}$ of the transform  coefficients with the qWPs $\left\{\Psi_{\pm[m]}\right\}$ from the decomposition  level $m$ consist of $2^{m+1}$ blocks $\mathbf{Z}_{[m]}=\mathbf{Z}_{+[m]}\bigcup\mathbf{Z}_{-[m]},$ where $\mathbf{Z}_{\pm[m]}=\bigcup_{l=0}^{2^{m}-1}\mathbf{Z}_{\pm[m],l}$. The blocks $\mathbf{Z}_{\pm[m],l}$ are  related to the  qWPs $\left\{\Psi_{\pm[m],l}\right\},\;l=0,...,2^{m}-1$, respectively.
The qWP transform  coefficients of a signal  $\mathbf{x}=\left\{x[k]\right\}\in\Pi[N]$ are the inner products of \textbf{x} with the translations of the corresponding wavelet packets:
\begin{equation}\label{coeLD1}
\begin{array}{ll}
   z_{\pm[m],l}[k]  =& \left\langle \mathbf{x},\Psi_{\pm[m],l}[\cdot-2^{m}k]\right\rangle=\sum_{j=0}^{N-1}\Psi^{*}_{\pm[m],l}[j-2^{m}k]\,x[j] =y_{[m],l}[k]\mp i\,c_{[m],l}[k],\\
 y_{[m],l}[k]  = &  \left\langle \mathbf{x},\psi_{[m],l}[\cdot-2^{m}k]\right\rangle,\quad c_{[m],l}[k]=\left\langle \mathbf{x},\phi_{[m],l}[\cdot-2^{m}k]\right\rangle,\quad l=0,...,2^{m}-1.
\end{array}
 \end{equation}
The transform s are implemented in a multiresolution mode by   multirate filtering. It follows from Proposition \ref{pro:cwq_des} that the structure of the filter bank for the transform  of a signal  \textbf{x} to the first decomposition  level differs from the structure of the filter banks for subsequent levels. Define the filter banks  $\mathbf{H}=\left\{ \mathbf{h}_{s}\right\}, \mathbf{F}=\left\{ \mathbf{f}_{s}\right\}, \mathbf{Q}_{\pm}=\left\{ \mathbf{q}_{\pm s}\right\}, \;s=0,1,$ by their impulse responses:
  \begin{equation}\label{hfq}
  h_{s}[k]=\psi_{[1],s}[k],\quad f_{s}[k]=\phi_{[1],s}[k],\quad q_{\pm s}[k]=h_{s}[k]\pm i f_{s}[k], \quad k=0,...,N-1,\;s=0,1.
\end{equation}
The four blocks
  $\left\{\mathbf{Z}_{+[1],0},\mathbf{Z}_{+[1]1},\mathbf{Z}_{-[1]0},\mathbf{Z}_{-[1]1}\right\}$ of the  first-level transform  coefficients  are derived by filtering the signal  \textbf{x} by the time reversed filters
  $\mathbf{Q}^{*}_{\pm s}=\mathbf{Q}_{\mp s}$, which is  followed by downsampling,  such as $ z_{\pm[1],l}[k]  =\sum_{j=0}^{N-1}q_{\mp\la}[j-2k]\,x[j], \;l=0,1.$

\begin{figure}
\centering
\includegraphics[width=3in]{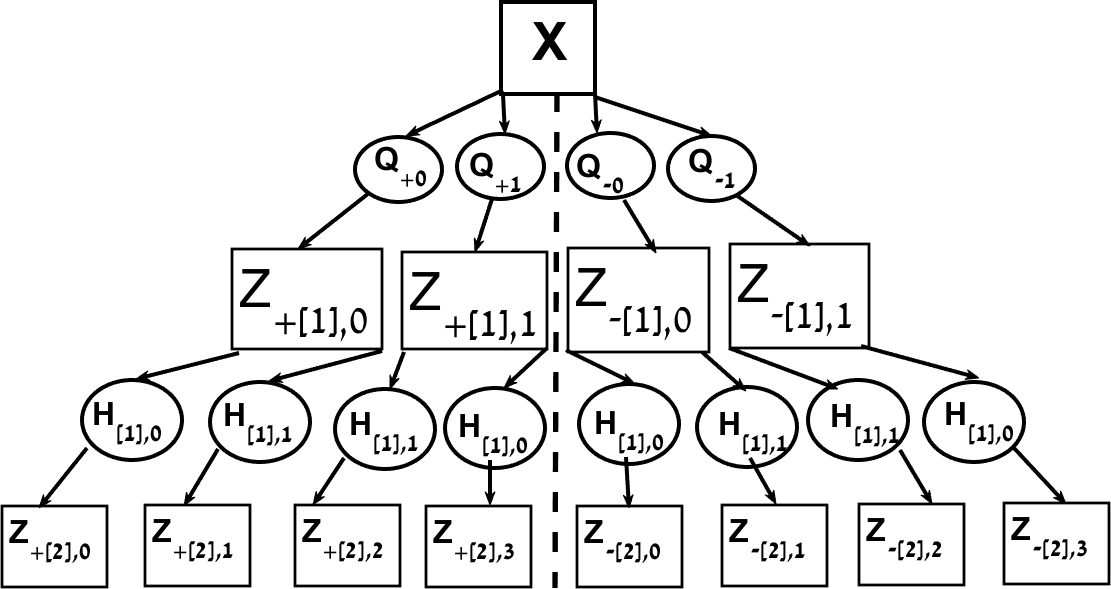}\hfill \vline \hfill\includegraphics[width=3in]{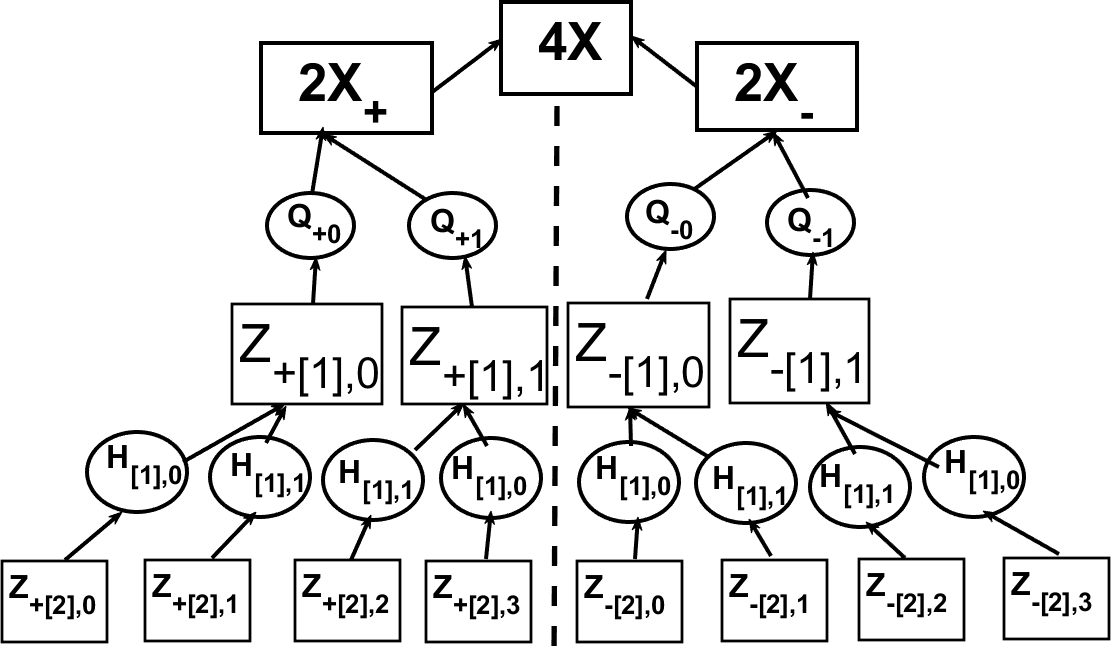}
\caption{Diagrams of the two-level 1D qWP transforms. Left: direct \t. Right: inverse \t}
\label{dia_trans}
\end{figure}

 The frequency responses  of the  filters $\mathbf{h}_{s}$ are $\hat{h}_{s}[n]=\sum_{k=0}^{N-1}e^{2\pi ikn/N}h_{s}[k]=\hat{\psi}_{[1],s}[n],\;n=0,...,N-1,\;s=0,1$.
 The filters $\mathbf{h}_{[m],s},\;m=1,...,M,\,s=0,1,$ for the transform  from the first to the subsequent decomposition  levels are defined
  via their frequency responses: $\hat{h}_{[m],s}[n]=\hat{h}_{s}[2^{m}n],\;m=1,...,M,\,s=0,1$. Thus, the qWP transforms from the first to the second decomposition  level are
 \[\begin{array}{cc}
      z_{\pm[2],0}[k]  = & \sum_{l=0}^{N/2-1}h_{[1],0}[l-4 k]\, z_{\pm[1],0}[l],\quad z_{\pm[2],1}[k]  =  \sum_{l=0}^{N/2-1}h_{[1],1}[l-4 k]\, z_{\pm[1],0}[l], \\
     z_{\pm[2],2}[k]  = & \sum_{l=0}^{N/2-1}h_{[1],1}[l-4 k]\, z_{\pm[1],1}[l],\quad z_{\pm[2],3}[k]  = \sum_{l=0}^{N/2-1}h_{[1],0}[l-4 k]\, z_{\pm[1],1}[l],\quad .
   \end{array}
 \]
 The transforms to the subsequent decomposition  levels are executed similarly using the filters $\mathbf{H}_{[m],s},\;m=2,...,M,\,s=0,1$.
 The diagrams in Fig. \ref{dia_trans} illustrate the  qWP transform  implementation.

The inverse transforms from the level-$M$ coefficients $\mathbf{Z}_{\pm[M]}$ to the coefficients of the first decomposition  level  $\left\{\mathbf{Z}_{+[1],0},\mathbf{Z}_{+[1]1},\mathbf{Z}_{-[1]0},\mathbf{Z}_{-[1]1}\right\}$ are executed in a standard way by upsampling and filtering the upsampled arrays by the filters $\mathbf{H}_{[m],s},\;m=2,...,M,l=0,1$. Then, arrays $\mathbf{Z}_{+[1],0}$ and $\mathbf{Z}_{+[1]1}$ are upsampled and filtered by the filters  $\mathbf{Q}_{+ 0} $ and $\mathbf{Q}_{+ 1} $, respectively, thus producing a complex signal  $2\mathbf{x}_{+}$. Similarly, the signal  $2\mathbf{x}_{-}$ is derived from the coefficients $\mathbf{Z}_{-[1]l}$ using the filters  $\mathbf{Q}_{- s} ,\;s=0,1$.  The signal  \textbf{x} is restored by $\mathbf{x}=\mathfrak{Re}(\mathbf{x}_{+}+\mathbf{x}_{-})/4$.

Practically, the direct and inverse transforms are implemented in the frequency  domain using modulation matrices (see \cite{azn_pswq}).

The 2D transforms of an image \textbf{X} of size $N\times N$ are executed in a tensor product mode that means application of 1D transforms with the filters $\mathbf{Q}_{\pm s},\;s=0,1,$ to rows of  the image \textbf{X}, which is followed by application of the transforms with the filters $\mathbf{Q}_{+ s},\;s=0,1,$ to the columns of the derived coefficient  arrays. Thus the  eight  coefficient  arrays of size $N/2\times N/2$ are produced:  $\left\{\mathbf{Z}_{+[1],\la,\mu},\,\mathbf{Z}_{-[1],\la,\mu},\;\la,\mu=0,1\right\}$. Then,  the  coefficient  arrays are  processed in the same way but using the filters $\mathbf{P}_{[m],\la},\;m=2,...,M,\la=0,1$ instead of $\mathbf{Q}_{\pm\la},\;\la=0,1$. The arrays related to the ``positive" qWPs $\Psi_{++}$ and  arrays related to the ``negative" qWPs $\Psi_{+-}$ are processed separately forming a double-tree structure. At the second level we have 32 coefficient  arrays $\left\{\mathbf{Z}_{+[2],\la,\mu},\,\mathbf{Z}_{-[2],\la,\mu},\;\la,\mu=0,1,2,3\right\}$ of size $N/4\times N/4$  and so on.

The inverse transforms from the level-$M$ coefficients $\left\{\mathbf{Z}_{+[M],\la,\mu},\,\mathbf{Z}_{-[M],\la,\mu},\;\la,\mu=0,...,2^{M}-1\right\}$ to the coefficients of the first decomposition  level  $\left\{\mathbf{Z}_{+[1],\la,\mu},\,\mathbf{Z}_{-[1],\la,\mu},\;\la,\mu=0,1\right\}$ are executed in a standard tensor product way  using filters $\mathbf{H}_{[m],s},\;m=2,...,M,l=0,1$.
Then, the complex sub-images $\mathbf{X}_{+}$ and $\mathbf{X}_{-}$ are derived from the coefficients $\mathbf{Z}_{+[1],\la,\mu}$ and $\mathbf{Z}_{-[1],\la,\mu},\;\la,\mu=0,1,$
 by the filters  $\mathbf{Q}_{+ s} $ and $\mathbf{Q}_{-s},\;s=0,1 $, respectively.  Then, the signal  $\mathbf{X}\in\Pi[N,N]$ is restored by ${\mathbf{X}}=\mathfrak{Re}(\mathbf{X}_{+}+\mathbf{X}_{-})/8$.

\begin{figure}
\centering
\includegraphics[width=5.3in]{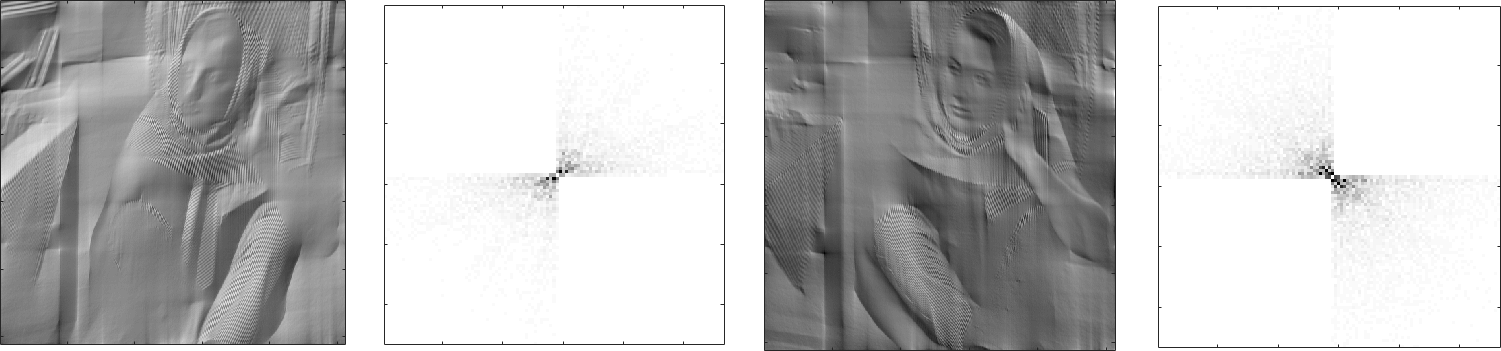}
\caption{Left to right: 1.Image  $\mathfrak{Re}(\mathbf{X}_{+})$. 2.Its magnitude DFT spectrum. 3.Image  $\mathfrak{Re}(\mathbf{X}_{-})$. 4.Its magnitude DFT spectrum }
\label{xp_xm_x5T2}
\end{figure}

Figure \ref{xp_xm_x5T2} illustrates   the ``Barbara" image restoration by the 2D signals $\mathfrak{Re}(\mathbf{X}_{\pm})$. The signal  $\mathfrak{Re}(\mathbf{X}_{+})$ captures edges oriented to \emph{north-east}, while $\mathfrak{Re}(\mathbf{X}_{-})$ captures edges oriented to \emph{north-west}. The signal  $\tilde{\mathbf{X}}=\mathfrak{Re}(\mathbf{X}_{+}+\mathbf{X}_{-})/8$ perfectly restores the image achieving  PSNR=313.8596 dB.

\section{Image inpainting methodology}\label{sec:s3}
\begin{table}
\resizebox{\textwidth}{!}{
  \centering
  \begin{tabular}{|l|l|}
     \hline
      \textbf{M1} -- Method1,  \textbf{M2} -- Method2,   \textbf{A1} -- \textbf{Algorithm I}, &$\Lambda_{1}[i]$ and $\Lambda_{2}[i]$ -- threshold sequences (\eh{La12})  \\
$ \mathbf{B}_{W[m]}^{\la}$  -- Bivariate shrinkage operator (\eh{bso}) & $\bar{\sigma}_{W[m]}[k,n]^{2}$ -- averaged variance for  $c_{[m]}[k,n]$ (\eh{barsi})  \\
   $\tilde{\sigma}_{W[m]}[k,n]^{2}$ -- marginal variance for $c_{[m]}[k,n]$ &$R_{1}$ and $R_{2}$ --free parameters  for thresholding\\
   $\mathbf{\tilde{F}}_{[m]}$ -- operator of  transform   $\mathbf{\tilde{F}}_{[m]}\,\mathbf{X}=\mathbf{Z}_{[m]}=\mathbf{Z}_{+[m]}\bigcup\mathbf{Z}_{-[m]}$& $\mathbf{F}_{[m]}$ -- reconstruction  operator from level $m$: $\mathbf{F}_{[m]}\,\mathbf{Z}_{[m]}=\mathbf{X}_{[m]}$\\
    $\Theta$ -- mask matrix,   $\Theta_{1}$ -- extended mask &$\rho=\frac{\sum_{l,n=1}^{N_{1}} \Theta_{1}[l,n]}{N_{1}^{2}}$, $tol_{1}$ and $tol_{2}$ --tolerance parameters\\
    \hline
   \end{tabular}
   }
  \caption{Notations for Section \ref{sec:s3}}\label{nota3}
\end{table}
 This section presents two qWP-based methods: Method1 (\textbf{M1}) and Method2 (\textbf{M2}) that are applied to image inpainting.
Image inpainting means restoration of an image that was degraded by having many missing pixels and, possibly, was  corrupted by Gaussian noise. Similarly to \textbf{Algorithm I}  (\textbf{AI}) in   \cite{braver,braver1}, both methods use decreasing thresholding values that are determined by the sequences $\Lambda_{1}[i]$ and $\Lambda_{2}[i]$ described in \eh{La12} and rely on the redundancy of the qWP transforms and on the interdependency between the transform  coefficients in the horizontal (between neighboring coefficients) and vertical (parent-child) directions. This interdependency is utilized via the Bivariate Shrinkage  algorithm  (BSA) (\cite{sen_seles}).

 \emph{We refer to the set of  the  filter banks  DAS-2,  DAS-1, TP-$\mathbb{C}$TF$_6$ and  TP-$\mathbb{C}$TF$_6^{\downarrow}$  as SET-4.}

 \textbf{AI}  uses the TP-$\mathbb{C}$TF$_6$ filter bank that produces excellent image inpainting results that outperform (in a PSNR sense) most state-of-the-art methods (see \cite{braver1} for methods comparison). In \cite{che_zhuang}, \textbf{AI}  is implemented using the SET-4  filter banks.
On the cartoon-type images such as ``Lena" and  ``Boat", TP-$\mathbb{C}$TF$_6$ achieves better performance than the rest of the members in SET-4. However,
      images with texture such as ``Fingerprint"  and, especially ``Barbara", are inpainted better by the filter bank DAS-2 with increased directionality compared to TP-$\mathbb{C}$TF$_6$. Unlike  \cite{che_zhuang,braver,braver1},  both \textbf{MI} and \textbf{M2} are  using 2D qWP transforms. The transform  coefficients from either two or three decomposition  levels are utilized.

\begin{description}
\item[Bivariate shrinkage operator (BSA):]
Assume that $c_{[m]}[k,n]$ is a transform  coefficient  from a decomposition  level $m$. The averaged variance for the coefficient  $c_{[m]}[k,n]$  is determined by the equation
\begin{equation}\label{barsi}
\bar{\sigma}_{W[m]}[k,n]^{2}=\frac{1}{(2W_{[m]})^{2}}\sum_{\k,\n=-W_{[m]}}^{W_{[m]}-1}c_{[m]}[k+\k,n+\n]^{2},
\end{equation}
where $W_{[m]}$ is an  integer.

The marginal variance for the  coefficient  $c_{[m]}[k,n]$ is estimated: $\tilde{\sigma}_{W[m]}[k,n]^{2}=(\bar{\sigma}_{W[m]}[k,n]^{2}-\la^{2})_{+},$ where $\la$ is a parameter taken from either the sequence $\Lambda_{1}$ or $\Lambda_{2}$ by the rule explained below.
Then, the BSA operator $\mathbf{B}_{W[m]}^{\la}$ is defined by
\begin{eqnarray}\label{bso}
  \mathbf{B}_{W[m]}^{\la}\,c_{[m]}[k,n]&\srr&\left\{
                          \begin{array}{ll}
                            c_{[m]}[k,n]-\la_{W[m]}(c)\,\frac{c_{[m]}[k,n]}{|c_{[m]}[k,n]|}, & \hbox{if $|c_{[m]}[k,n]|>\la_{W}(c)$;} \\
                            0, & \hbox{otherwise,}
                          \end{array}
                        \right. \\\nonumber  \la_{W[m]}(c)&=&\frac{\,\sqrt{3}\,\la^{2}}{\tilde{\sigma}_{W[m]}[k,n]\,\sqrt{1+|c_{[m+1]}[k,n]/c_{[m]}[k,n]|^{2}}},
\end{eqnarray}
where ${c}_{[m+1]}[k,n]$ is the transform  coefficient  from the coarser decomposition  level $m+1$ which is  located at approximately the same spatial position as the coefficient  $c_{[m]}[k,n]$. In our case, it is related to a waveform  with approximately the same directionality as the waveform  related to $c_{[m]}[k,n]$. If $\mathbf{C}_{m}=\left\{c_{[m]}[k,n]\right\}$ is the set of all transform  coefficients from the decomposition  level $m$, then $\mathbf{B}_{W}^{\la}\,\mathbf{C}_{m}$ is the application of the operator $\mathbf{B}_{W}^{\la}$ to each coefficient  $c_{[m]}[k,n]\in\mathbf{C}_{m}$.

  \item[Thresholding parameters $\la$: ] \label{tresh}
Assume that the image to be restored comprises $M$ pixels while  $K$ pixels are missing.  Denote by  $\rho=K/M$ the percentage of missing pixels.   Assume that the noise STD=$\sigma$ is known. Let $\la_{max}\srr 512,$
\begin{equation}\label{lamM}
  \la_{min}\srr\max\left\{1,\sigma\left(1-\frac{\rho^{2}}{2}\right)\right\}, \quad \la_{mid}\srr\min\left\{\max\left\{2\,\la_{min}+10\right\},20\right\}.
\end{equation}
Let $r_{1}\srr \la_{mid}/\la_{max}<1$, $r_{2}\srr \la_{min}/\la_{mid}<1$. The sequences  $\Lambda_{1}[i]$ and $\Lambda_{2}[i]$ are defined by
\begin{equation}\label{La12}
  \Lambda_{1}[j]=\sqrt{2}\,r_{1}^{\frac{j-R_{1}}{R_{1}-1}}\,\la_{mid},\;j=1,...,R_{1},\quad \Lambda_{2}[j]=\sqrt{2}\,r_{2}^{\frac{j-R_{2}}{R_{2}}}\,\la_{min},\;j=1,...,R_{2}
\end{equation}
where $R_{1}$ and $R_{2}$ are free parameters. In our experiments, we use $R_{1}=5$, $R_{2}=8$.

The parameters $\la$ in \eh{bso} are taken from the sequences $\Lambda_{1}[j]$ and $\Lambda_{2}[j]$.

      \item[Preliminaries common to  \textbf{M1} and   \textbf{M2}:] The problem to be solved is  to estimate  an  image $\mathbf{{X}}$ of size $N\times N$ from the degraded array $\mathbf{\check{X}}=\Theta\cdot(\mathbf{{X}}+\mathbf{n})$.  Here $\Theta$, which is referred to as the inpainting mask,  is an  $N\times N$ matrix consisting of ones and zeros and \textbf{n} is zero-mean Gaussian noise with STD=$\sigma$.  We assume that the mask   $\Theta$ and $\sigma$ are known.

Denote by
$\mathbf{\tilde{F}}^{+}_{[m]}$ and $\mathbf{\tilde{F}}_{-[m]}$ the  operators of the  qWP transforms  of an  image $\mathbf{X}$  to the coefficients of decomposition  level $m$ such that   $\mathbf{\tilde{F}}_{\pm[m]}\,\mathbf{X}=\mathbf{Z}_{\pm[m]}$.  The transforms with the complex qWPs $\Psi^{p}_{++[m],j,l}$ and $\Psi^{p}_{+-[m],j,l}$, respectively.  Denote by $\mathbf{\tilde{F}}_{[m]}$ the combined operator such that $\mathbf{\tilde{F}}_{[m]}\,\mathbf{X}=\mathbf{Z}_{[m]}=\mathbf{Z}_{+[m]}\bigcup\mathbf{Z}_{-[m]}$.
Denote by $\mathbf{{F}}_{+[m]}$ and $\mathbf{{F}}_{-[m]}$ the  operators  of  the  reconstruction  of images $\mathbf{X}_{\pm[m]}$
from the sets of the $m$-level transform  coefficients $\mathbf{Z}{+[m]}$ and $\mathbf{Z}_{-[m]}$, respectively: $\mathbf{X}_{\pm[m]}=\mathbf{F}_{\pm[m]}\mathbf{Z}{\pm[m]}$. Denote by $\mathbf{F}_{[m]}$ the operator of the  full reconstruction  of the image from the coefficient  array $\mathbf{Z}_{[m]}=\mathbf{Z}_{+[m]}\bigcup\mathbf{Z}_{-[m]}$. It means that
$\mathbf{F}_{[m]}\,\mathbf{Z}_{[m]}=\mathbf{X}_{[m]}=\mathfrak{Re}(\mathbf{X}_{+[m]}+\mathbf{X}_{-[m]})/8.
$ The solution is obtained by a weighted average of reconstruction  results from several levels such as  $\tilde{\mathbf{X}}=\frac{\sum_{m}\b_{m}\mathbf{X}_{[m]}}{\sum_{m}\b_{m}}$, where either $m=3,4$ or $m=2,3,4$.

  \item[Settings common to  \textbf{M1} and   \textbf{M2}:]The following free parameters' values are set in advance: 1. Order $p$ of the DTS that generates qWPs (in most cases  $p=3,\,4$ or 5). 2. Integers $R_{1}$ and  $R_{2}$ (\eh{La12}) and tolerance parameters $tol_{1}$ and $tol_{2}$ (typically, $tol_{1}=0.05$ and $tol_{2}=0.01$). 3. Iterations limits $L_{1}$, $L_{2}$ and  $L_{3}$; 4. Windows spans $\{W_{m}\}$   for the averaged variances' $\bar{\sigma}_{W[m]}[k,n]$ calculation  (\eh{barsi}) and  balance weights $\{\b_{m}\}$,  where $m=(2,)3,4.\;$

 In order to eliminate boundary effects, the degraded image $\mathbf{\check{X}}$ of size $N\times N$ is symmetrically extended to the  image $\mathbf{Y}$ of size $N_{1}\times N_{1}$, where $N_{1}=N+2T$. Typically, $T=N/8$. Respectively, the mask $\Theta$ is extended to  $\Theta_{1}$.
  \end{description}
\subsection{Method1}\label{sec:ss31}
Assume that $m=3,4$. Extension to $m=2,3,4$ is straightforward.
This method is a slight modification of  \textbf{AI}.  The solution scheme is based on the assumption that the original
image $\mathbf{X}$  can be
sparsely represented in the  qWP transform  domain.

\textbf{AI} (and   \textbf{M1}) consist of iterative thresholding of the transform  coefficients with decreasing localized thresholds. After setting a few parameters, the  threshold levels are determined automatically by using BSA. 

 \underline{ Initialization: } Let 
$  \mathbf{X}^{0}=\mathbf{X}^{1}=0$,  $k=\n=1$.  Compute the parameter $\rho=\frac{\sum_{l,n=1}^{N_{1}} \Theta_{1}[l,n]}{N_{1}^{2}}$ and sequences $ \Lambda_{1}$ and $ \Lambda_{2}$ (Eqs. \rf{lamM} and \rf{La12}).
The thresholding parameter is $\la=\Lambda_{1}[1]$. Then, the iterations have the following steps:
\begin{enumerate}
\item $\mathbf{Y}^{k}=\Theta_{1}\cdot Y+(\mathbf{X}^{k}-\Theta_{1}\cdot \mathbf{X}^{k})$;
\item Compute $\mathbf{Z}^{k}_{\pm[m]}=\mathbf{\tilde{F}}_{\pm[m]}\,\mathbf{Y}^{k},\;m=3,4,5$\footnote{The coefficients $\mathbf{Z}^{k}_{\pm[5]}$ from the fifth decomposition  level are used in the BSA  operators.};
\item Select the thresholding parameter $\la$/(stop the iterations) by the \textbf{Select/stop rule} (below);
\item Update the arrays $\mathbf{Z}^{k}_{\pm [m]}$ by  the application of  BSA operators $\mathbf{B}_{W{[m]}}^{\la},\;m=3,4,$ to these arrays:
\(\;\mathbf{Z}^{k+1}_{\pm [m]}= \mathbf{B}_{W{[m]}}^{\la}\,\mathbf{Z}^{k}_{\pm [m]},\quad m=3,4; \)
\item Apply the inverse qWP transforms to the arrays $\mathbf{Z}^{k+1}_{ [m]}$, where $m=3,4,$
  \[ \mathbf{X}^{k+1}_{[m]}=\mathbf{F}_{[m]}\,\mathbf{Z}^{k+1}_{ [m]},\quad \mathbf{X}^{k+1}\srr\frac{\b_{3}\mathbf{X}^{k}_{[3]}+\b_{4}\mathbf{X}^{k}_{[4]}}{\b_{3}+\b_{4}};\]
 \item  $k=k+1$. Go to 1.
\end{enumerate}

\paragraph{Select/stop rule common to \textbf{M1} and \textbf{M2}:}\label{ssru}The rule, to some extent,  is similar to the rule  in \textbf{AI} in   \cite{braver1}.  Iterations limits are $L_{1}$, $L_{2}$ and  $L_{3}$ (not present in \textbf{AI}). Initial thresholding parameter is $\la=\Lambda_{1}[1]$ and $\n=1$. Denote the number of iterations with a certain index $\n$ by $K_{\n}$.
\begin{itemize}
  \item The $l_2$ norm of the difference $\Delta^{k}=\left\| \mathbf{X}^{k}-\mathbf{X}^{k-1}\right\|_{2}$ is computed;
  \item If $\n<R_{1}  \hbox{ \textbf{and} }\left\{ K_{\n}>L_{1}  \hbox{ \textbf{or} }\Delta^{k}<tol_{1} \right\}$, then $\n=\n+1$ and  $\la=\Lambda_{1}[\n]$;
  \item If $R_{1}\leq \n<R_{1}+R_{2} \hbox{ \textbf{and} } \left\{ K_{\n}>L_{2} \hbox{ \textbf{or} }\Delta^{k}<tol_{2}\right\}$, then $\n=\n+1$ and  $\la=\Lambda_{2}[\n-R_{1}]$;
\item If $\n=R_{1}+R_{2} \hbox{ \textbf{and} }\left\{ K_{\n}>L_{3} \hbox{ \textbf{or} } \Delta^{k}<tol_{2}\right\}$, then \textbf{STOP} iterations;
\item The array $\mathbf{X}^{k}$ is shrunk  to the original size $N\times N$ as $\mathbf{X}^{k}\succ \mathbf{\tilde{X}}$, and the array $\mathbf{\tilde{X}}$ is taken as a solution to the inpainting problem.
\end{itemize}
\subsection{Method2}\label{sec:ss32}
An  image restoration scheme is developed in Chapter 18 in \cite{ANZ_book1} by utilizing non-directional 2D  wavelet  frames designed in  Chapter 17  in \cite{ANZ_book1}.  Images are restored in \cite{ANZ_book1} by the application of the \emph{split
Bregman iteration} (SBI) scheme \cite{gold_os} that uses the
so-called \emph{\aa-based} approach (see for example \cite{ji_shen_xu}).

\textbf{M2}  to be described couples \textbf{M1}  with  the SBI scheme. In essence, it is the  SBI algorithm  that uses  qWPs and is supplied with  decreasing localized thresholds determined by BSA and the \emph{Select/stop rule}. In addition to the settings that are common to \textbf{M1} and \textbf{M2}, we set the regularization parameter $\mu.$

  \underline{ Initialization:}  Let $k=\n=1$,
$ \mathbf{X}^{0}=0$,  $\mathbf{b}^{0}_{+[m]}=\mathbf{b}^{0}_{-[m]}=0$,  $\mathbf{b}^{0}_{[m]}=\mathbf{b}^{0}_{+[m]}\bigcup\mathbf{b}^{0}_{-[m]}$,    $\mathbf{d}^{0}_{+[m]}=\mathbf{d}^{0}_{-[m]}=0$, $\mathbf{d}^{0}_{[m]}=\mathbf{d}^{0}_{+[m]}\bigcup\mathbf{d}^{0}_{-[m]}$,  $m=3,4,5$\footnote{The coefficients $\mathbf{d}^{k}_{[5]},\;\mathbf{b}^{k}_{[5]}$ from the fifth decomposition  level are used in the BSA  operators.}.
 Compute the parameter $\rho=\frac{\sum_{l,n=1}^{N_{1}} \Theta_{1}[l,n]}{N_{1}^{2}}$ and sequences $ \Lambda_{1}$ and $ \Lambda_{2}$ (Eqs. \rf{lamM} and \rf{La12}). The thresholding parameter is $\la=\Lambda_{1}[1]$. Then, the iterations have the following steps:
\begin{enumerate}
\item Apply the inverse qWP transforms to the arrays $\mathbf{D}_{[m]}^{k-1}=\mathbf{d}^{k-1}_{[m]}-\mathbf{b}^{k-1}_{[m]}$, where $m=3,4,5$:
  \[ \left\{\mathbf{x}^{k-1}_{[m]}\srr\mathbf{F}_{[m]}\,\mathbf{D}_{[m]}^{k-1},\right\},\quad \mathbf{x}^{k-1}\srr\frac{\b_{3}\mathbf{x}^{k-1}_{[3]}+\b_{4}\mathbf{x}^{k-1}_{[4]}}{\b_{3}+\b_{4}};\]
  \item  The next iteration $\mathbf{X}^{k}$ is derived from \eh{breg_1} that is solved by the \emph{conjugate gradient} algorithm:  \begin{equation}\label{breg_1}
    \mathbf{X}^{k}:=\Theta_{1}\cdot\mathbf{x} +\mu\,\mathbf{x}=\mathbf{Y}+\mu\,\mathbf{x}^{k-1};
\end{equation}
\item Select the thresholding parameter $\la$/(stop the iterations) by the \textbf{Select/stop rule} ( above).
  \item Apply the forward qWP transforms to the array $\mathbf{X}^{k}$:
\(\mathbf{\tilde{F}}^{\pm}_{[m]}\,\mathbf{\mathbf{X}^{k}}=\mathbf{Z}^{k}_{\pm[m]}, \) and denote $\tilde{\mathbf{Z}}^{k}_{\pm[m]}=\mathbf{Z}^{k}_{\pm[m]}+\mathbf{b}^{k-1}_{\pm [m]}$, where  $m=3,4,5$;

\item Update the arrays $\mathbf{d}^{k-1}_{\pm [m]}$ by application of the  BSA operators $\mathbf{B}_{W{[m]}}^{\la},\;m=3,4,$ to the arrays $\tilde{\mathbf{Z}}^{k}_{\pm[m]}$:
\(\;\mathbf{d}^{k}_{\pm [m]}= \mathbf{B}_{W{[m]}}^{\la}\,\tilde{\mathbf{Z}}^{k}_{\pm[m]},\quad  \mathbf{d}^{k}_{[m]}=\mathbf{d}^{k}_{+[m]}\bigcup\mathbf{d}^{k}_{-[m]},\quad m=3,4,5; \)
\item Update the arrays $\mathbf{b}^{k-1}_{\pm [m]}$ by $\mathbf{b}^{k}_{\pm [m]}=\mathbf{b}^{k-1}_{\pm [m]}+\left(\tilde{\mathbf{Z}}^{k}_{\pm[m]}-\mathbf{d}^{k}_{\pm [m]} \right),\quad  \mathbf{b}^{k}_{[m]}=\mathbf{b}^{k}_{+[m]}\bigcup\mathbf{b}^{k}_{-[m]},\quad m=3,4,5$;
\item  $k=k+1$. Go to item 1.
\end{enumerate}

\section{Experimental results}\label{sec:s4}
\begin{figure}
\centering
\includegraphics[width=3.75in]{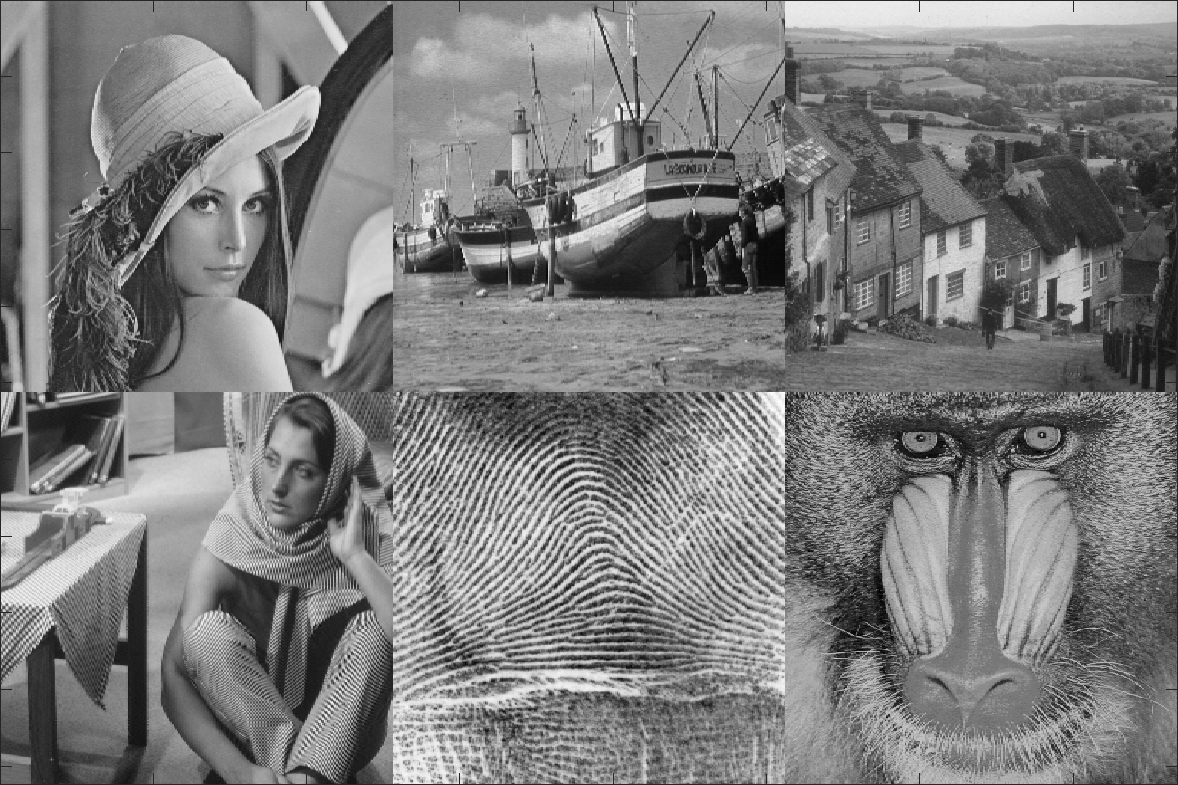}\hfill\includegraphics[width=2.5in]{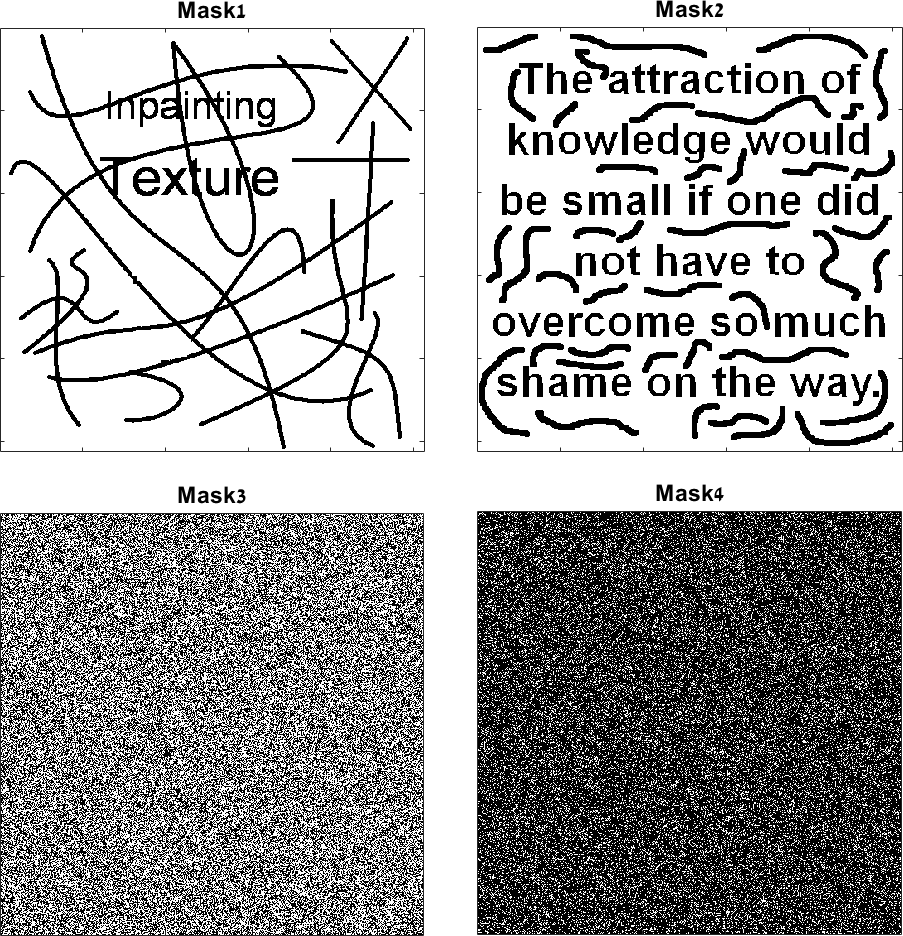}
 \caption{Left: Clean images: ``Lena",   ``Boat",   ``Hill",    ``Barbara",    ``Fingerprint"   and  ``Mandrill". Right: Four types of masks: Mask1 and Mask2 -- text and  curves; Mask3 and  Mask4-- 50\% and  80\% of random pixels missing, respectively}
\label{clima}
\end{figure}

For the experiments, we use a standard set of benchmark images, which was used in \cite{che_zhuang}: ``Lena",   ``Boat",    ``Barbara",  and    ``Fingerprint"   with two additional images ``Hill" and    ``Mandrill". The ``clean" images are displayed in Fig. \ref{clima}.
The performances of   \textbf{M1} and \textbf{M2} are compared with the performances of the state-of-the-art inpainting  algorithms that use the  SET-4 filter banks. The inpainting results  by the SET-4  algorithms applied to the first four images from Fig. \ref{clima}, which are degraded by the  application of four masks displayed in Fig. \ref{clima} and additive Gaussian  noise with various intensities ( STD $\sigma=0, 5, 10, 30, 50$ dB), are presented in \cite{che_zhuang}. The Matlab codes, which produce the results in \cite{che_zhuang}, are available at \\ http://staffweb1.cityu.edu.hk/xzhuang7/softs/index.html\#bdTPCTF. We used these codes to  restore the additional images ``Hill" and    ``Mandrill".  For each triple \emph{image-mask-$\sigma$}, the inpainting results from the application of  our \textbf{M1} and \textbf{M2} are compared with  the best result produced by the application of the SET-4  algorithms. The results are compared according to  PSNR and SSIM values and by visual inspection.
\br\label{rem19_20}SSIM values for the restored images are computed by using the Matlab function  \texttt{ssim}. It should be noted that the \texttt{ssim} function  in the Matlab 2020a version produces higher SSIM values than Matlab 2019b. The SSIM results listed below are computed using Matlab 2019b.\er

\subsection{Experiments with ``Barbara" and  ``Fingerprint"   images}\label{sec:sss41}
The evaluated images have complicated structure which comprises smooth regions, multiple edges (in ``Barbara" ) and oscillatory  patterns (in both images). The restoration results in PSNR and SSIM measurements  are given in Table \ref{tabbarfinI}.

\begin{table}
 \centering
\resizebox{\textwidth}{!}{%
\begin{tabular}{|l|l|l|l|l|l|l}
    \hline
    \multicolumn{4}{|c|}{Barbara} &\multicolumn{3}{|c|}{Fingerprint}\\\hline
    $\sigma$ &\textbf{M1}& \textbf{M2}& Best  from SET-4 &\textbf{M1}& \textbf{M2}& Best from SET-4 \\\hline
    \multicolumn{7}{|c|}{mask1}\\
    \hline
    0 & 38.13/0.958 & \textbf{38.42/0.967} & {36.68}/0.958 (4 )& 31.93/0.978& \textbf{32.64/0.980} &31.72/0.976 (4)\\\hline%
   5 & {34.93}/0.8256 & \textbf{34.94/0.831}&34.05/0.8138(1)& 30.69/0.963 & \textbf{30.86/0.964}&30.36/0.960 (1)\\\hline
    10 & \textbf{32.57}/0.748 &32.56\textbf{/0.753}& 32.02/0.739 (1)&28.96/0.939 & \textbf{29.09/0.940}&28.77/0.935 (1) \\\hline
    30 & \textbf{28.09}/0.601 & 28.01\textbf{/0.603}& 27.74/0.592 (1)& 25.15/0.843  &  \textbf{25.25/0.852}&25.04/0.839(1)\\\hline%
    50 & \textbf{25.85}/ 0.501 & 25.77\textbf{/0.504}& 25.51/0.489 (1)& 23.26/0.767 & \textbf{23.42/0.778}&23.21/0.766 (1)\\\hline
     \multicolumn{7}{|c|}{mask2}\\
    \hline
    0 & 34.26/0.922 & \textbf{34.70/0.932} & 33.66/0.923 (1 )& 28.76/0.949& \textbf{28.94/0.950} &28.31/0.942 (4)\\\hline
     5 & \textbf{32.78}/0.80 & 32.75 \textbf{/0.806}&32.24/0.787 (1)  & 28.02/\textbf{0.938}& \textbf{28.31}/0.936&27.68/0.926(1)\\\hline
   10 & \textbf{31.13}/0.730&31.11 \textbf{/0.734}& 30.71/0.718 (1)&27.01/0.908 & \textbf{27.08/0.911}&26,76/0.903 (1)\\\hline%
    30 & \textbf{27.22}/0.579& 27.1\textbf{/0.581}& 26.95/0.570 (1)& 24.16/0.812 &  \textbf{24.35/0.822}&24.15/0.809 (1)\\\hline%
   50 & \textbf{25.21}/0.487 & 25.13\textbf{/0.487} & 24.96/0.471 (1) & 22.58/0.739  & \textbf{22.77/0.753} & 22.65/0.738 (1)\\\hline
    \multicolumn{7}{|c|}{mask3}\\
        \hline
     0 & 37.30/0.901 & \textbf{37.48/0.913} & 35.72/0.9051 (1 )& 34.53/0.979&\textbf{34.54/0.979} &34.19/0.977 (4)\\\hline
    5 & \textbf{34.24}/0.789 & 34.21\textbf{/ 0.793}&33.29/0.778 (1)&31.72/0.958 &\textbf{ 31.92/0.961}&31.54/0.957 (4)\\\hline
    10 & 31.82/0.721& \textbf{31.85/0.723}&31.12/0.707 (1) &29.14/0.926  & \textbf{29.37/0.932}&29.09/0.927 (4)  \\\hline
    30 & 27.1/0.571 & \textbf{27.17/0.572}&26.77/0.554 (1)& 24.51/0.809 &  \textbf{24.73/0.825}&24.43/0.806 (4)\\\hline%
    50 & {24.84}/0.460& \textbf{24.9/0.467}&24.60/0.441 (1)&22.48/0.721 & \textbf{22.69/0.748}&22.53/0.722 (1)\\\hline
    \multicolumn{7}{|c|}{mask4}\\
        \hline
      0 &30.17/0.7609  & \textbf{30.34/0.779} & 29.12/0.753 (1 )&\textbf{27.03}/0.899 & 27.02/0.9 &26.77/0.894 (4)\\\hline%
    5 & 29.28/0.687  &  \textbf{29.44/0.696}&28.41/0.673 (1)&{26.27/0.878}  &\textbf{26.29/0.881} &25.87/0.872(4)\\\hline
    10 & 28.01/0.626& \textbf{28.19/0.633}&27.25/0.609 (1) & 25.11/0.843 &\textbf{25.11/0.845} &24.60/0.834 (4) \\\hline
    30 & \textbf{24.59}/0.452 & 24.45\textbf{/0.461}&24.23/0.434(1)& 22.09/0.712  & \textbf{22.15/0.731}&22.11/0.7115 (1)\\\hline
    50 &\textbf{22.67}/0.347& 22.58\textbf{/0.365}&22.35/325 (1)&  20.32/0.613&20.31/\textbf{0.651}&\textbf{20.61}/0.614 (1)\\\hline
\end{tabular}\vline
}
  \caption{PSNR/SSIM values for the restoration of ``Barbara" (columns 2--4) and  ``Fingerprint" (columns 5--7) images by \textbf{M1}, \textbf{M2} and the  best algorithm from SET-4.
    Boldface highlights  the best results. $\sigma$ -- noise STD. Numbers in parentheses indicate which is the best  algorithm from SET-4  that produces the best result: (1) means DAS-2, (2)--DAS-1, (3)--TP-$\mathbb{C}$TF$_6$, (4)--TP-$\mathbb{C}$TF$_6^{\downarrow}$
   }\label{tabbarfinI}
\end{table}

\begin{figure}
\centering
\includegraphics[width=3.in]{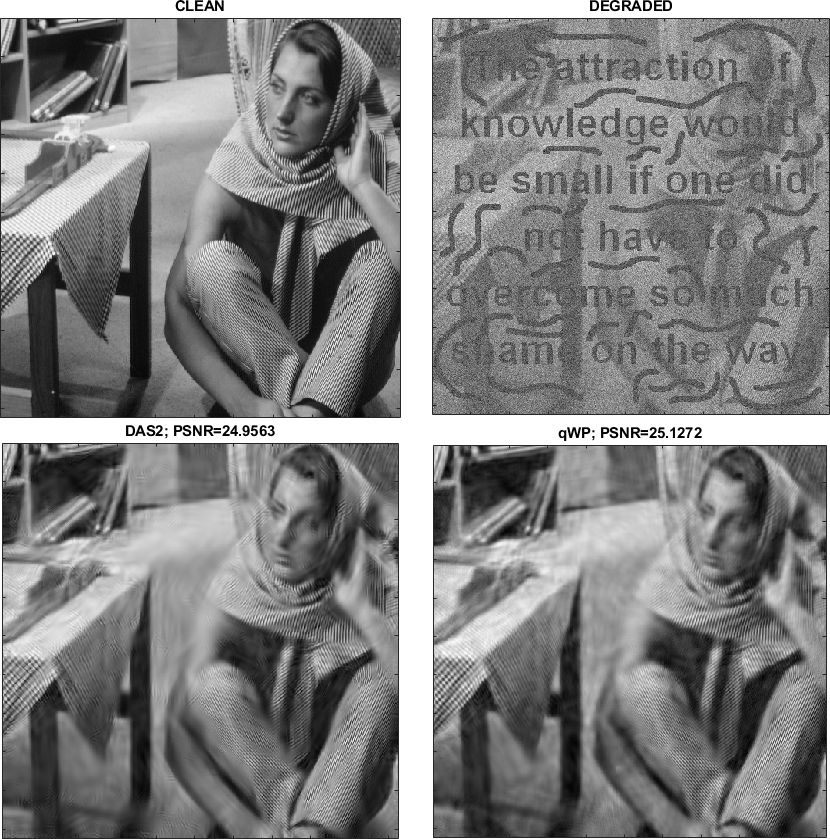}\hfill\includegraphics[width=3.2in]{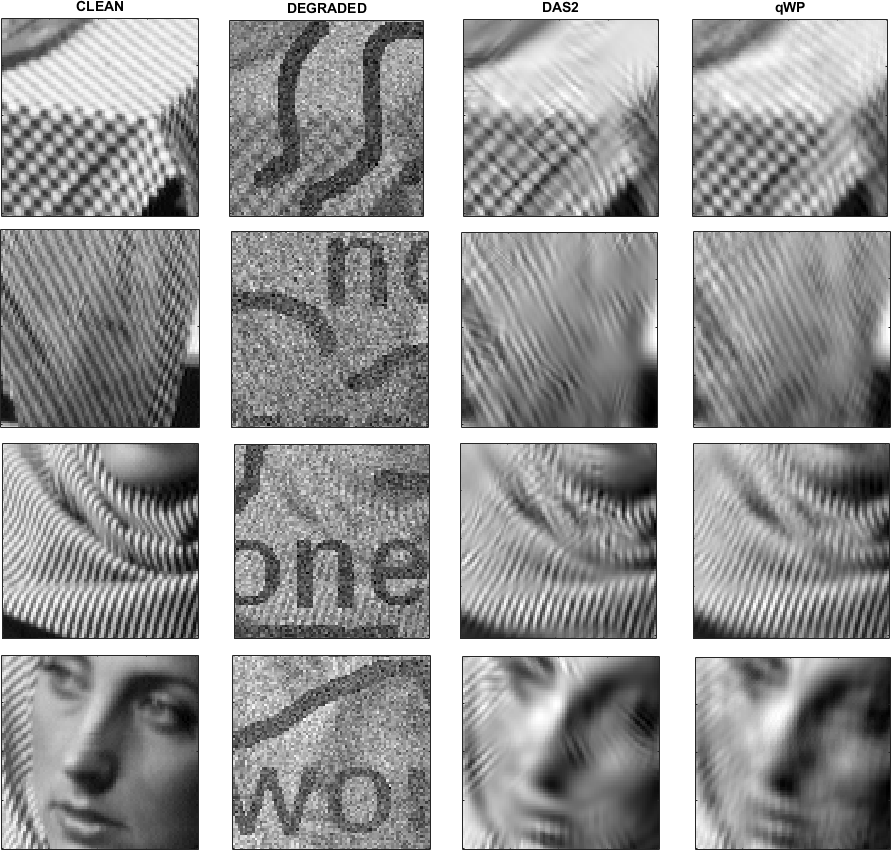}
\caption{Restoration of ``Barbara" image. \textbf{Left:} Top left: clean image. Top right: image degraded by application to it mask2 and   Gaussian noise with $\sigma=50$ dB. Bottom: Restoration by \textbf{M2} (right), PSNR=25.13 dB,  SSIM=0.4868 and by  DAS-2  (left), PSNR=24.96 dB, SSIM=0.4709. \textbf{Right:} Fragments of    images. Columns: First - clean fragments; Second -- degraded; Third--   restored by  DAS-2; Fourth -- restored by \textbf{M2}}
\label{barb2_50_4}
\end{figure}

Figure \ref{barb2_50_4}  illustrates  the restoration of the  ``Barbara"  image, which was   degraded by the application of mask2 and by adding a strong    Gaussian noise with $\sigma=50$ dB, by  \textbf{M2} and DAS-2. Both algorithms successfully inpainted the mask and suppressed the noise but, in doing so, DAS-2 loses fine texture details and adds some artifacts into the image. Figure \ref{barb2_50_4} demonstrates it.

\begin{figure}
\centering
\includegraphics[width=3.in]{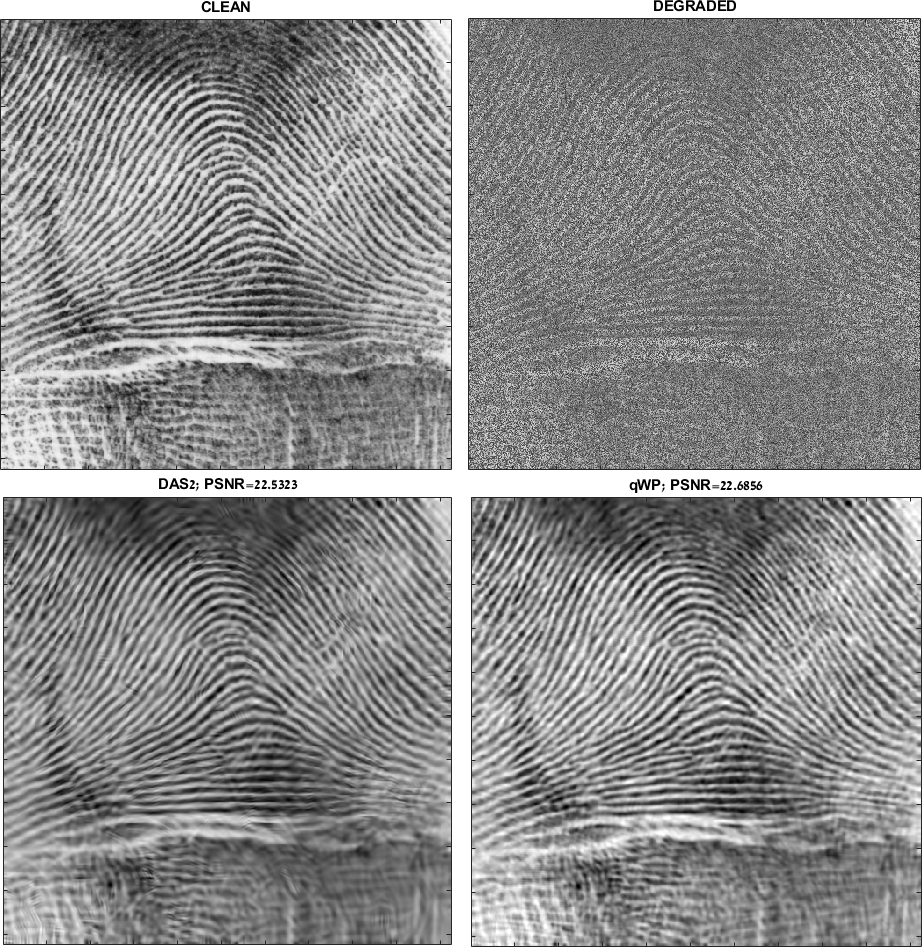}\hfill \includegraphics[width=3.2in]{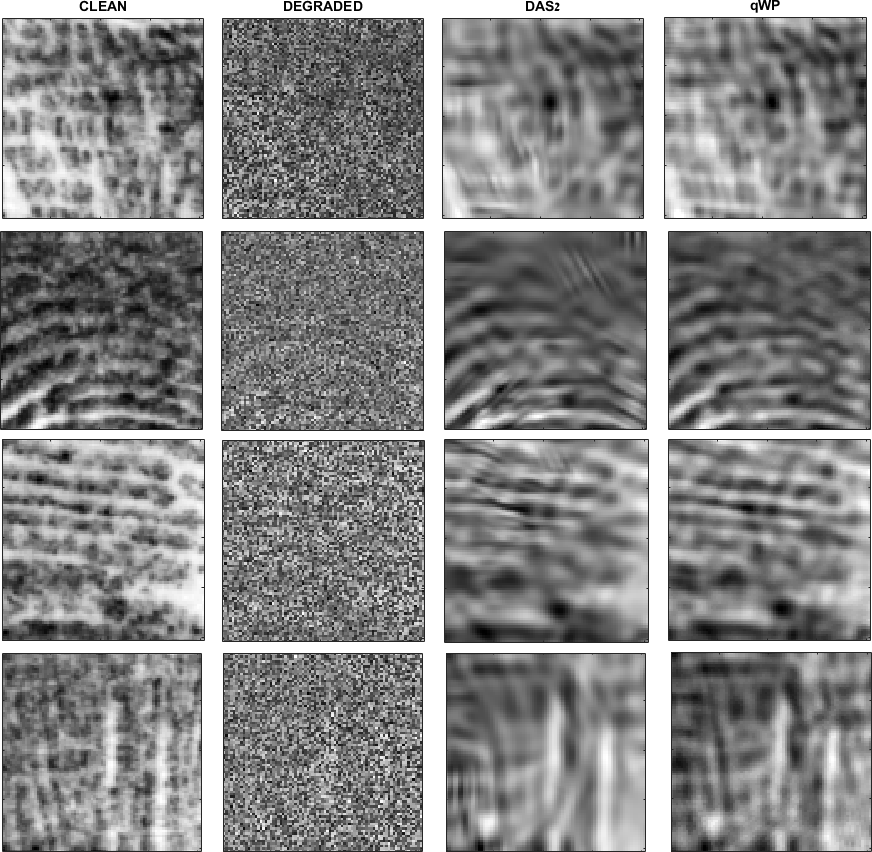}
\caption{ \textbf{Left:} Restoration of ``Fingerprint" image. Top left: clean image. Top right: image degraded by application to it  mask3 and   Gaussian noise with $\sigma=50$ dB.
Bottom: Restoration by \textbf{M2} (right), PSNR=22.69 dB, SSIM=0.748 and by  DAS-2  (left), PSNR=22.53 dB, SSIM=0.7224. \textbf{Right:} Fragments of    images. Columns: First - clean fragments; Second -- degraded; Third--   restored by  DAS-2; Fourth -- restored by \textbf{M2}}
\label{fing3_50_4}
\end{figure}

                                                 Figure \ref{fing3_50_4}  illustrates the  restoration of the ``Fingerprint" image, which was severely  degraded by the application of mask3 (50\% of the pixels are missing) and strong additive   Gaussian noise with $\sigma=50$ dB, by  \textbf{M2} and DAS-2. In this  case, the PSNR and  SSIM for restoration by  \textbf{M2} are higher than those for  the restoration by DAS-2. Both algorithms successfully inpainte the mask and suppress the noise, but DAS-2  over-smooths some parts of  the image. It is seen in Fig. \ref{fing3_50_4}.

\subsection{Experiments with ``Lena" and ``Boat" images}\label{sec:ss42}
These images are characterized by a number of edges separating smooth areas and some fine details  such as  hair and the hat decoration in ``Lena" and the ground in  ``Boat". The PSNR and SSIM results are given in Table \ref{tablenboaI}.

\begin{table}
\centering
\resizebox{\textwidth}{!}{%
\begin{tabular}{|l|l|l|l|l|l|l}
    \hline
    \multicolumn{4}{|c|}{Lena} &\multicolumn{3}{|c|}{Boat}\\\hline
    $\sigma$ &\textbf{M1}& \textbf{M2}& Best from SET-4 &\textbf{M1}& \textbf{M2}& Best from SET-4 \\\hline
    \multicolumn{7}{|c|}{mask1}\\
    \hline
    0 & 37.50/0.9371 & 37.84/\textbf{0.9488} & \textbf{38.02}/0.9487 (3) & 34.78/0.936 & 34.95/\textbf{0.94072} & \textbf{34.96}/0.94071 (3)\\\hline%
    5 & 34.94/0.702 & 34.95/\textbf{0.724}&\textbf{35.19}/0.672 (3)& 32.68/0.778 & 32.68/\textbf{0.792}&\textbf{32.81}/0.752 (3)\\\hline
    10 & 33.18/0.605  &33.22/\textbf{0.614}&\textbf{ 33.42}/0.586 (3)& 30.79/0.659  &30.75/\textbf{0.674}&\textbf{31.04}/0.630 (3) \\\hline
    30 &29.34/0.449 & 29.43/\textbf{0.454}& \textbf{29.81}/0.449 (3)&27.03/0.454 &27.02/\textbf{0.463}& \textbf{27.41}/0.433 (3)\\\hline%
    50 & 27.31/0.370 & 27.40/\textbf{0.377}& \textbf{27.85}/0.377 (3) & 25.22/0.348 & 25.15/\textbf{0.360}& \textbf{25.6}/0.334 (2)\\\hline
      \hline
       \multicolumn{7}{|c|}{mask2}\\
          \hline
    0 & 33.72/0.884  & 33.96/\textbf{0.9005} & \textbf{34.3}/0.9001 (3)& 30.23/0.874  & 30.57/0.880 & \textbf{30.80/0.881} (3)\\\hline%
    5 & 32.48/0.661& 32.43/\textbf{0.692} &\textbf{32.97}/0.647 (3)& 29.30/0.721& 29.29/\textbf{0.741}&\textbf{29.79}/0.718 (1)\\\hline
    10 & 31.39/0.582  &31.37/\textbf{0.593}&\textbf{31.78}/0.567 (3) & 28.18/0.611 &28.25/\textbf{0.630}&\textbf{28.85}/0.59 (2) \\\hline
    30 &28.39/0.432 & 28.51/\textbf{0.440}& \textbf{28.89}/0.434 (3)&25.78/0.417 &25.83/\textbf{0.430}&\textbf{26.34}/0.407 (2)\\\hline%
    50 & 26.68/0.356 & 26.65/\textbf{0.364}&\textbf{27.22}/0.362 (3)& 24.33/0.319& 24.37/\textbf{0.334}& \textbf{24.84}/0.312 (2)\\\hline
      \hline
       \multicolumn{7}{|c|}{mask3}\\
         \hline
  0 & 37.70/0.813   & 37.95/\textbf{0.8528} & 38.0/0.852 (3)& 34.39/0.8518  & \textbf{34.47/0.865} & 34.42/0.858 (3)\\\hline
    5 &35.14/0.647& 35.17/\textbf{0.660 }&\textbf{35.40}/0.632 (3)& 32.26/0.700& 32.29\textbf{/0.731}&\textbf{32.50}/0.684 (3)\\\hline%
    10 & 33.03/0.568 &33.04/\textbf{0.577}&\textbf{33.40}/0.558 (3) & 30.3/0.598 &30.41/\textbf{0.623}&\textbf{30.65}/0.586 (3) \\\hline
    30 &28.67/0.418 & 28.81/\textbf{0.426}& \textbf{29.18}/0.421 (3)&26.33/0.405  &26.32/\textbf{0.428}&\textbf{26.66}/0.380 (3)\\\hline%
    50 & 26.51/0.341 & 26.65/\textbf{0.349}&\textbf{27.06}/0.346 (3)& 24.43/ 0.299& 24.49/\textbf{0.335}&\textbf{24.75}/0.278 (3)\\\hline
      \hline
       \multicolumn{7}{|c|}{mask4}\\
    \hline
   0 & 32.09/0.6311  & 32.16/\textbf{0.6756} &32.33/0.6740 (3)& 28.39/0.629 & 28.25/\textbf{0.644} & \textbf{28.58}/0.643 (3)\\\hline%
    5 &31.18/0.552& 31.15/ \textbf{0.553}&\textbf{31.44}/0.551 (3)& 27.7/0.544& 27.71/\textbf{0.558}&\textbf{27.98}/0.546 (3)\\\hline%
    10 & 29.86/0.484 &29.84/\textbf{0.4874}&\textbf{30.25} /0.487(3) & 26.77/0.465&26.73/\textbf{0.480} &\textbf{27.08}/0.465 (3) \\\hline
    30 &26.40/0.3406 & 26.52/\textbf{0.349}& \textbf{26.95}/0.347 (3)&24.01/0.278&24.0/\textbf{0.308} &\textbf{24.46}/0.277 (3)\\\hline%
    50 &24.25/0.260 &24.47/\textbf{0.2753}&\textbf{25.15}/0.2746 (3) &22.50/0.196&22.22/\textbf{0.233}&\textbf{22.96}/0.193 (3)\\\hline
  \end{tabular}\vline
  }
  \caption{PSNR/SSIM values for the restoration of ``Lena" (columns 2--4) and  ``Boat" (columns 5--7) images by \textbf{M1}, \textbf{M2} and the  best  from SET-4  algorithms.
    Boldface highlights  the best results. $\sigma$ -- noise STD. Numbers in parentheses indicate which algorithm in  SET-4  produces the best result: (1) means DAS-2, (2)--DAS-1, (3)--TP-$\mathbb{C}$TF$_6$, (4)--TP-$\mathbb{C}$TF$_6^{\downarrow}$
   }
  \label{tablenboaI}
\end{table}

\begin{figure}
\centering
\includegraphics[width=3.in]{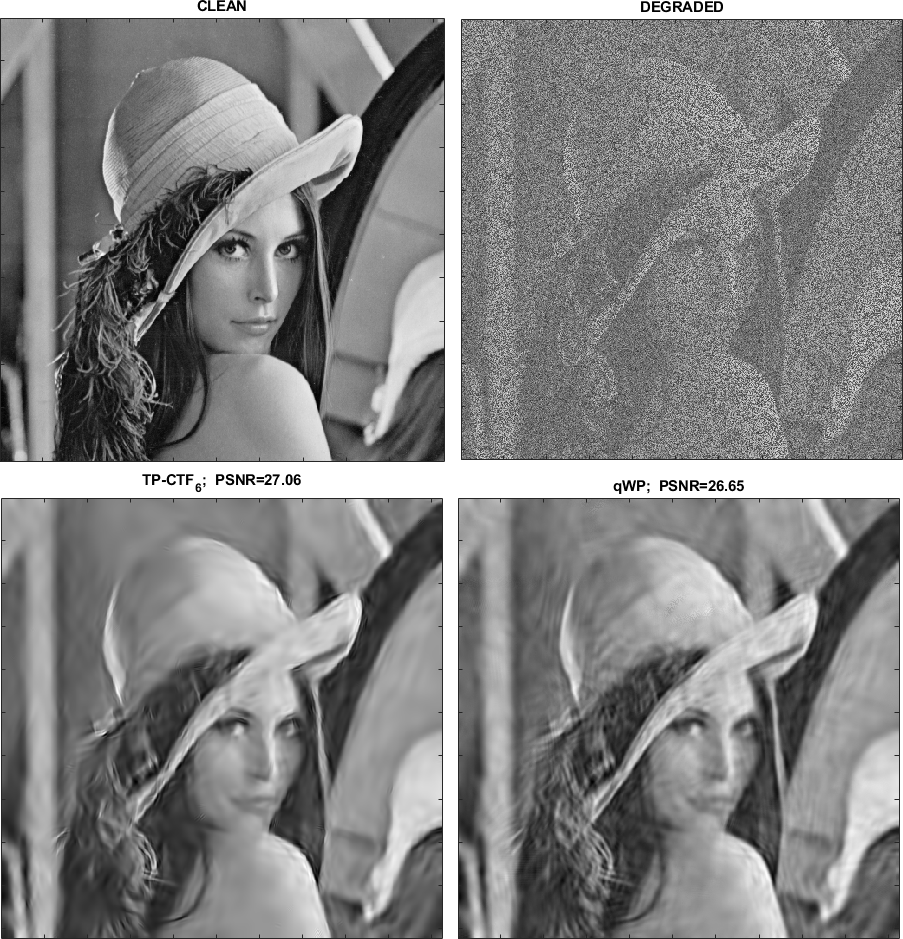}\hfill
\includegraphics[width=3.2in]{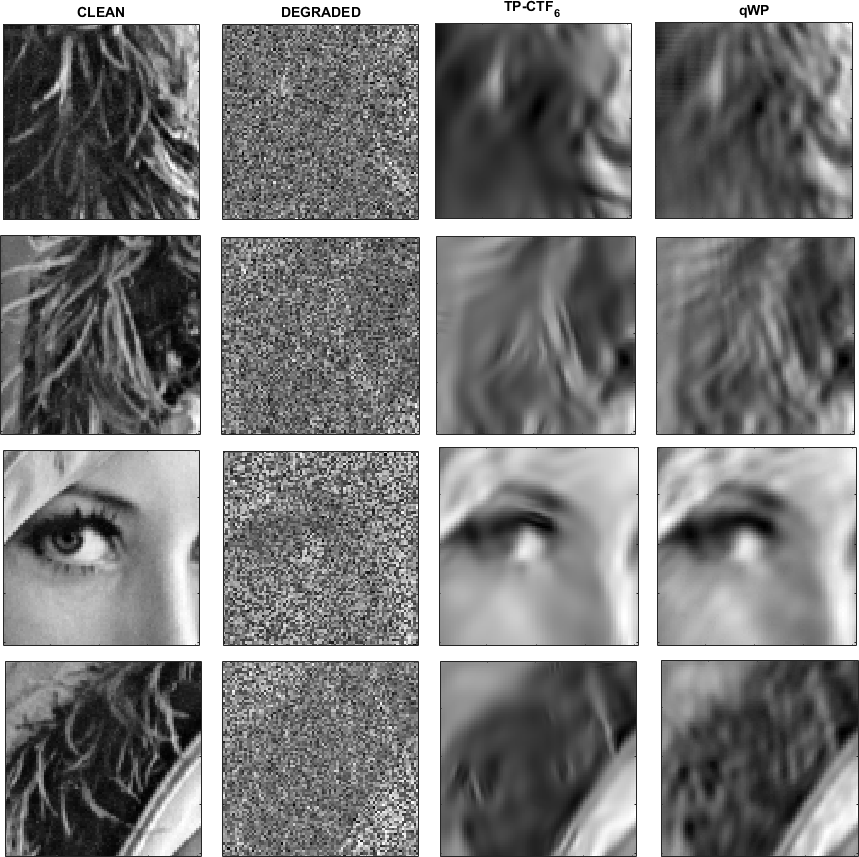}
\caption{Restoration of  the ``Lena" image.  \textbf{Left:}Top left: clean image. Top right: image degraded by application of mask3 and   Gaussian noise with $\sigma=50$ dB.
Bottom: Restoration by \textbf{M2} (right), PSNR=26.65 dB, SSIM=0.3494 and by  TP-$\mathbb{C}$TF$_6$   (left), PSNR=27.05 dB, SSIM=0.3457. \textbf{Right:} Fragments of    images. Columns: First - clean fragments; Second -- degraded; Third--   restored by  DAS-2; Fourth -- restored by \textbf{M2}}
\label{lena3_50_4}
\end{figure}

Figure \ref{lena3_50_4}  displays the restoration  of the ``Lena" image, which was severely  degraded by the application of mask3 (50\% of pixels are missing), and strong Gaussian noise with $\sigma=50$ dB, by  \textbf{M2} and TP-$\mathbb{C}$TF$_6$, which was the best from the SET-4 algorithms. The PSNR of the restoration by  TP-$\mathbb{C}$TF$_6$ is higher than what \textbf{M2} has produced. On the other hand,  the SSIM values are close to each other while some fine details, which are lost by TP-$\mathbb{C}$TF$_6$, are retained by \textbf{M2}.

\begin{figure}
\centering
\includegraphics[width=3.in]{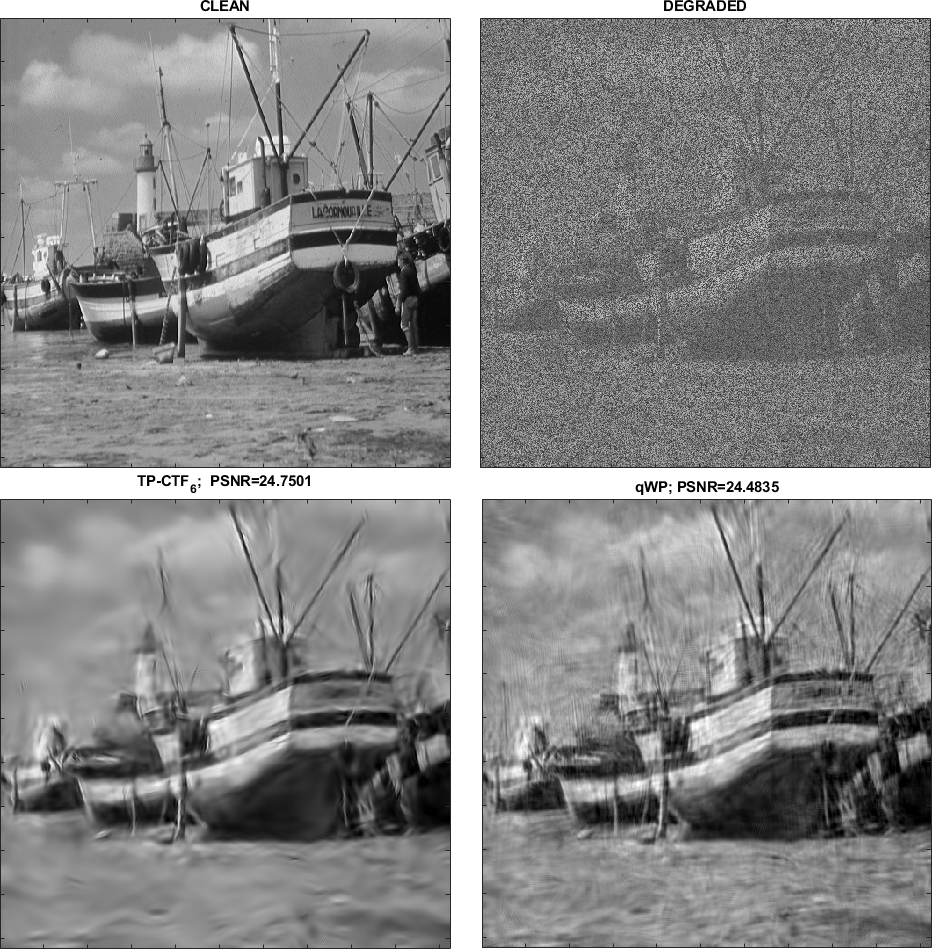}\hfill
\includegraphics[width=3.2in]{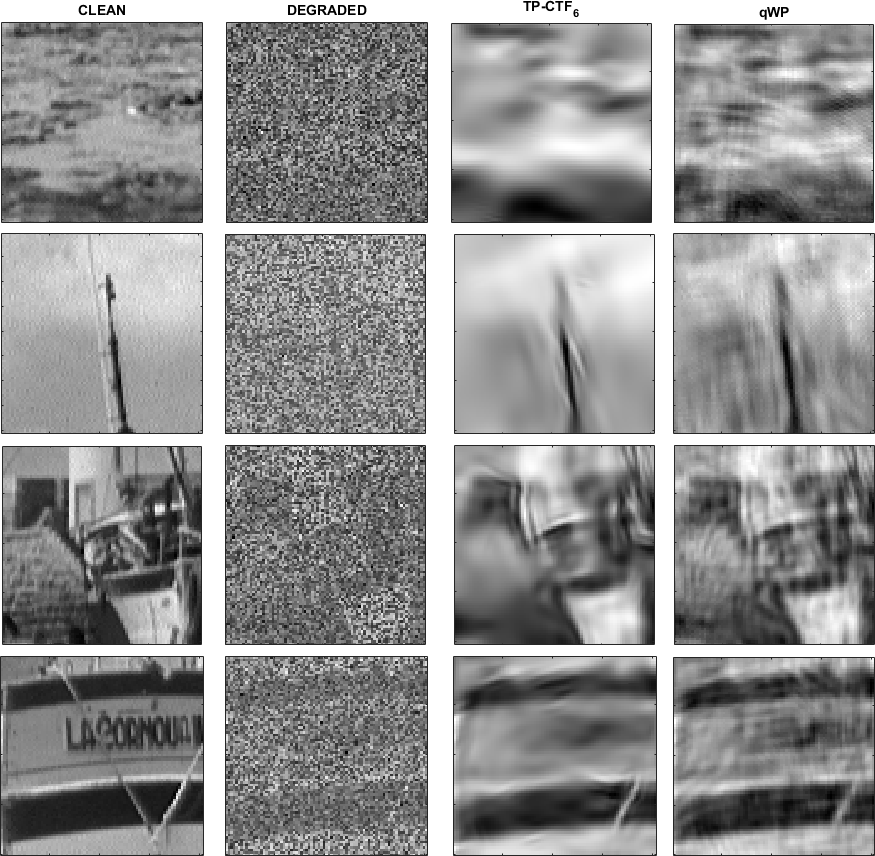}
\caption{Restoration of ``Boat" image.  \textbf{Left:}Top left: clean image. Top right: image degraded by application of mask3 and   Gaussian noise with $\sigma=50$ dB.
Bottom: Restoration by \textbf{M2} (right), PSNR=24.48 dB, SSIM=0.3347 and by  TP-$\mathbb{C}$TF$_6$   (left), PSNR=24.75 dB, SSIM=0.2777. \textbf{Right:} Fragments of    images. Columns: First - clean fragments; Second -- degraded; Third--   restored by  DAS-2; Fourth -- restored by \textbf{M2}}
\label{boat3_50_4}
\end{figure}

 Figure \ref{boat3_50_4} displays the results of the ``Boat" restoration by  \textbf{M2} and TP-$\mathbb{C}$TF$_6$, which produced the highest PSNR from the SET-4 algorithms.  The image was  degraded by the application of mask3 (50\% of pixels are missing) and strong additive  Gaussian noise with $\sigma=50$ dB.   Although the PSNR for restoration by  TP-$\mathbb{C}$TF$_6$ is higher than what  \textbf{M2} has produced,   the \textbf{M2}-SSIM value is much higher compared to TP-$\mathbb{C}$TF$_6$.  The  image   restored by TP-$\mathbb{C}$TF$_6$  looks  over-smoothed compared to restoration by \textbf{M2}. Some edges and almost the entire texture are lost by the application of TP-$\mathbb{C}$TF$_6$ but are  retained by \textbf{M2}.

\subsection{Additional examples: ``Hill"  and  ``Mandrill"}\label{sec:ss43}
We conducted experiments with the images ``Hill"  and  ``Mandrill" that were not used in  \cite{che_zhuang}. The former image is smooth up to the texture on the roofs and ground, whereas the most part of the latter one consists of very fine texture. The performance of   \textbf{M2} is compared with the performances of  algorithms that use the  SET-4 filter banks. The inpainting results  by the SET-4  algorithms are derived using the Matlab codes at the web site\\ http://staffweb1.cityu.edu.hk/xzhuang7/softs/index.html\#bdTPCTF. The PSNR and SSIM results are given in Table \ref{tablehill}. In most experiments with the ``Hill" image, the PSNR values produced by SET-4 algorithms are higher than the values produced by \textbf{M2} but for the SSIM values the situation is quite opposite. In many experiments, the \textbf{M2}-SSIM values are much higher compared to those from SET-4. For the   ``Mandrill" image, all the SSIM and almost all the PSNR values produced by the SET-4 algorithms are inferior to those produced by  \textbf{M2}.

\begin{table}
\centering
\resizebox{9cm}{!}
{
\begin{tabular}{|l|l|l|l|l}
    \hline
    \multicolumn{3}{|c|}{Hill} &\multicolumn{2}{|c|}{Mandrill}\\\hline
    $\sigma$ & \textbf{M2}& Best from SET-4 & \textbf{M2}& Best from SET-4 \\\hline
    \multicolumn{5}{|c|}{mask1}\\
    \hline
    0  & \textbf{36.2/0.9419} & {35.97}/0.9403 (3)  & \textbf{30.13/0.9291} & 30.02/0.9278  (1)\\\hline%
    5 & \textbf{33.55/0.8082}&{33.49}/0.7851 (3)& \textbf{28.99/0.8678} & 28.89/0.8609 (1)\\\hline
    10  &\textbf{31.53/0.7033}&{ 31.41}/0.6816 (1)&\textbf{ 27.32/0.7864}  &27.22/0.7667 \\\hline
    30 & 27.81/\textbf{0.4774}& \textbf{27.89}/0.4556 (1)&\textbf{23.57/0.5634} &23.51/0.5333\\\hline%
    50 & 25.99/\textbf{0.3683}& \textbf{26.19}/0.3473 (1) &\textbf{ 21.89/0.412 }& 21.86/{0.3875}\\\hline
      \hline
       \multicolumn{5}{|c|}{mask2}\\
          \hline
    0   & 32.44/\textbf{0.8884} & \textbf{32.74}/0.8876 (3)& \textbf{27.43/0.8652}  & 27.37/0.8638 (1) \\\hline%
    5  &31.11/\textbf{0.765} &\textbf{31.37}/0.7465 (1)& \textbf{26.76/0.805}& 26.74/{0.8}(1)\\\hline
    10 & 29.79/\textbf{0.6618}&\textbf{30.03}/0.6434 (1) & \textbf{25.76/0.7325 }&25.71/{0.7106} (1) \\\hline
    30 & 26.78/\textbf{0.4404}& \textbf{27.15}/0.4245 (1)&\textbf{22.97/0.5238} &22.96/{0.4897} (1)\\\hline%
    50 &  25.38/\textbf{0.3427}&\textbf{25.69}/0.3229 (1)& \textbf{21.5601/0.3843}& 21.558/{0.3495}(2)\\\hline
      \hline
       \multicolumn{5}{|c|}{mask3}\\
         \hline
  0    & \textbf{34.69/0.8674} & 34.52.0/0.8667 (3)& \textbf{26.44/ 0.8276}  & {25.92/0.8127} (1)\\\hline
    5  &32.57/\textbf{0.7529 }&\textbf{32.59}/0.7326 (3)&\textbf{25.88/0.7768}& 25.17/{0.7413} (1)\\\hline%
    10 & 30.64/\textbf{0.6522}&\textbf{30.7}/0.6077 (3) & \textbf{25/0.7028}&24.42/{0.6653} (1) \\\hline
    30  & {26.87}/\textbf{0.4369}&27.14/0.3948 (1)   &\textbf{22.39/0.5017}& 22.2/0.441 (1)\\\hline%
    50  & 25.2/\textbf{0.3245}&\textbf{25.4}/0.26 (1)& \textbf{21.05/ 0.355}& 20.99/{0.2887} (1)\\\hline
      \hline
       \multicolumn{5}{|c|}{mask4}\\
    \hline
   0 &  29.34/\textbf{0.6756} &29.59/0.6659 (3)& \textbf{21.84/0.5644} & 21.48/{0.5431}  (1)\\\hline%
    5 & 28.58/ \textbf{0.5846}&\textbf{28.87}/0.5736 (3)&\textbf{ 21.54/0.5328}& 20.93/{0.4951} (1)\\\hline%
    10  &27.59/\textbf{0.4978}&\textbf{17.91} /0.4731(3) & \textbf{21.17/0.4773}&20.67/{0.4455}  (1) \\\hline
    30  & 24.67/\textbf{0.3103}& \textbf{25.35}/0.2935 (1)&19.8/\textbf{0.3264}&\textbf{20.23}/{0.2977}  (1)\\\hline%
    50&23.33/\textbf{0.2753}&\textbf{24.06}/0.2094 (1) &18.93/\textbf{0.2323}&\textbf{19.81}/0.1414  (1)\\\hline
  \end{tabular}\vline
  }
  \caption{PSNR/SSIM values for the restoration of ``Hill" (columns 2--3) and  ``Mandrill" (columns 4--5) images by \textbf{M2} and the  best  from SET-4  algorithms.
    Boldface highlights  the best results. $\sigma$ -- noise STD. Numbers in parentheses indicate which algorithm in  SET-4  produces the best result: (1) means DAS-2, (3)--TP-$\mathbb{C}$TF$_6$
   }
  \label{tablehill}
\end{table}

\begin{figure}
\centering
\includegraphics[width=3.in]{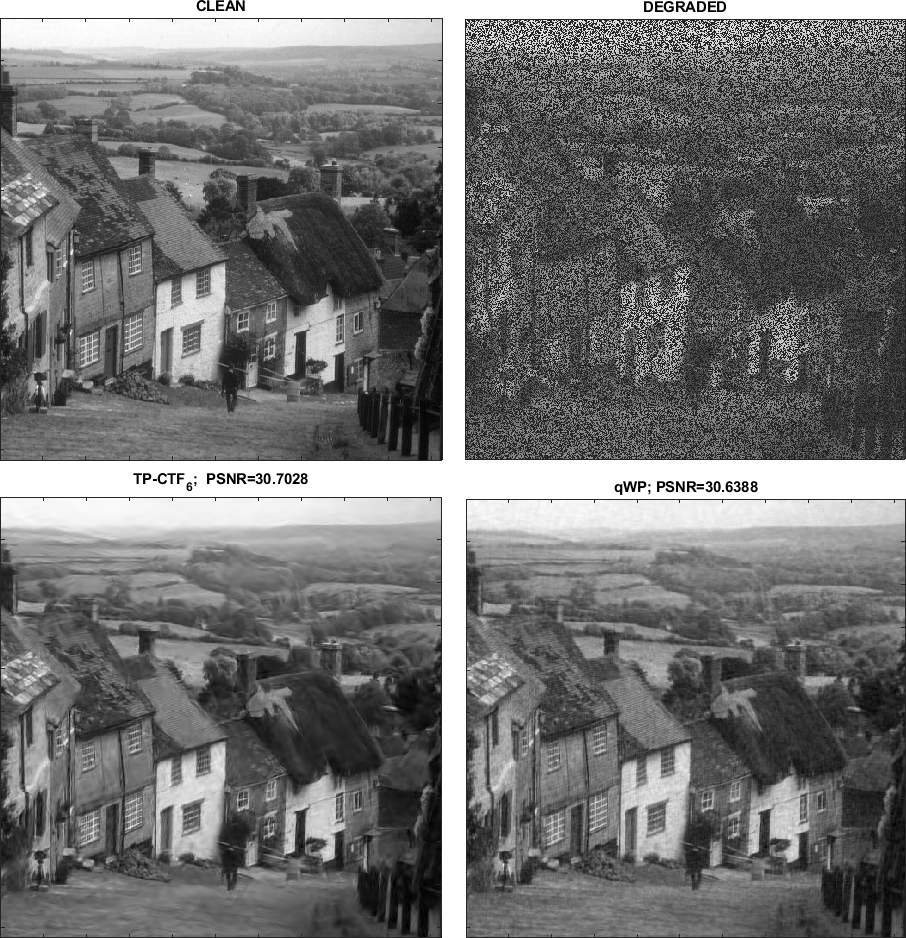}\hfill
\includegraphics[width=3.in]{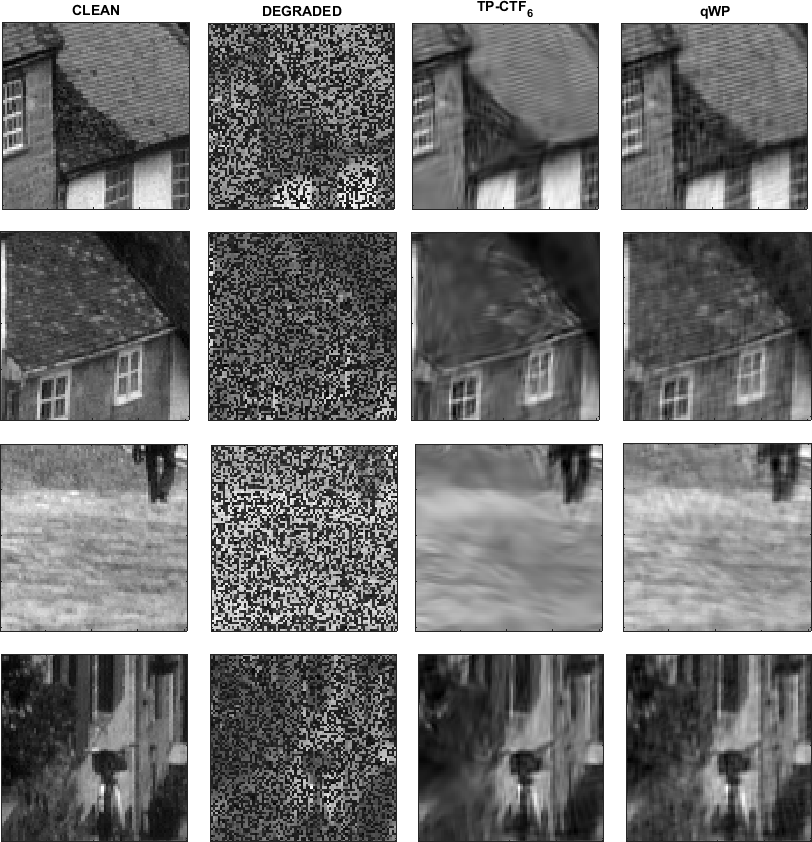}
\caption{Restoration of ``Hill" image.  \textbf{Left:}Top left: clean image. Top right: image degraded by application of mask3 and   Gaussian noise with $\sigma=10$ dB.
Bottom: Restoration by \textbf{M2} (right), PSNR=30.64 dB, SSIM=0.6522 and by  TP-$\mathbb{C}$TF$_6$   (left), PSNR=30.7 dB, SSIM=0.6077. \textbf{Right:} Fragments of    images. Columns: First - clean fragments; Second -- degraded; Third--   restored by  TP-$\mathbb{C}$TF$_6$; Fourth -- restored by \textbf{M2}}
\label{hill3_10}
\end{figure}

Figure \ref{hill3_10} displays the restoration results of the ``Hill" image, which was   degraded by the application of mask3 (50\% of pixels are missing), and moderate Gaussian noise with $\sigma=10$ dB, by  \textbf{M2} and TP-$\mathbb{C}$TF$_6$, which has produced the highest PSNR out of the SET-4 algorithms. The PSNR of the restoration by  TP-$\mathbb{C}$TF$_6$ is higher than that from \textbf{M2}. On the other hand,  the SSIM value from the \textbf{M2} restoration  significantly exceeds that from TP-$\mathbb{C}$TF$_6$. Consequently,  fine details, which are lost by TP-$\mathbb{C}$TF$_6$, are retained by \textbf{M2}.

\begin{figure}
\centering
\includegraphics[width=3.in]{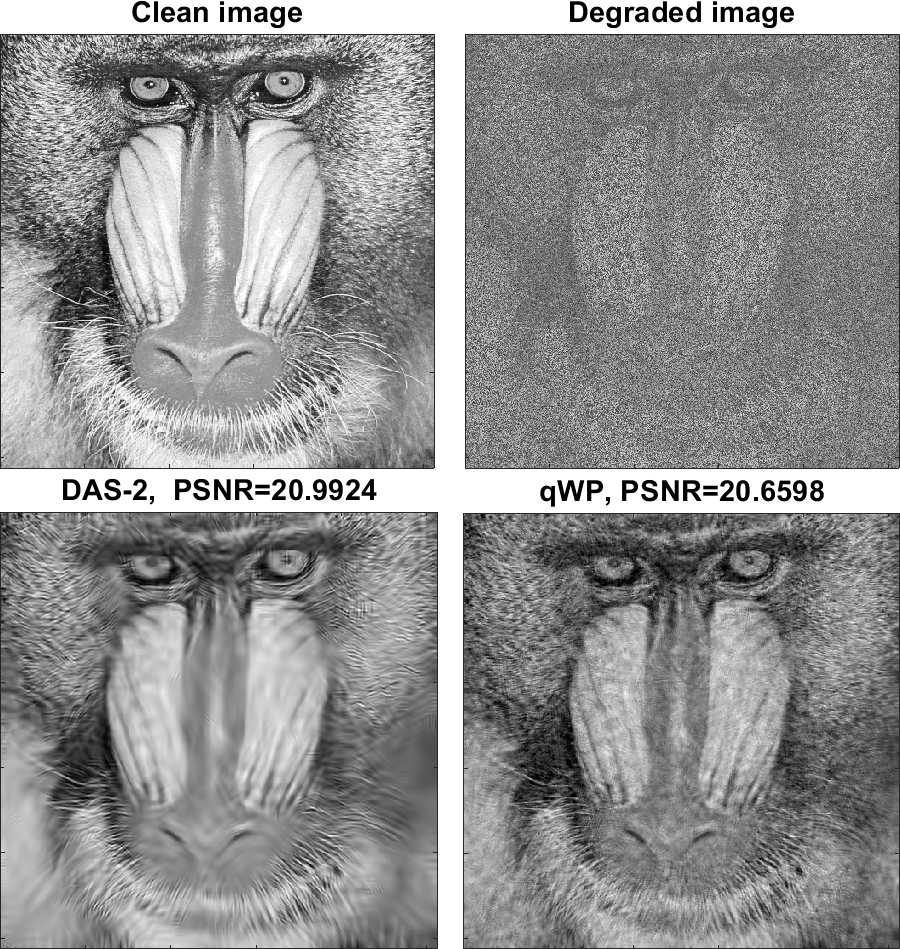}\hfill
\includegraphics[width=3.in]{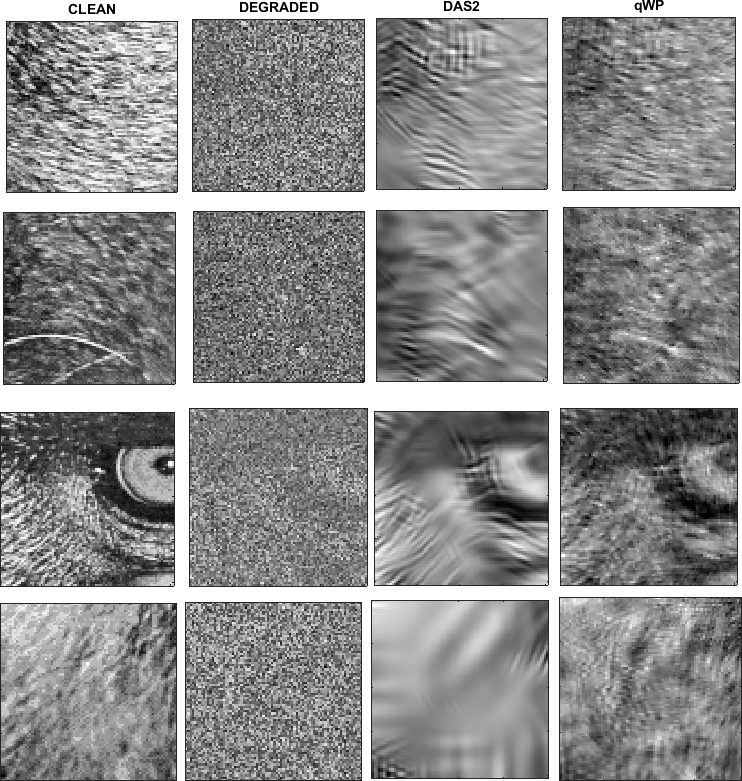}
\caption{Restoration of  the ``Mandrill" image.  \textbf{Left:}Top left: clean image. Top right: image degraded by application of mask3 and   Gaussian noise with $\sigma=50$ dB.
Bottom: Restoration by \textbf{M2} (right), PSNR=21.05 dB, SSIM=0.355 and by  DAS-2   (left), PSNR=20.99 dB, SSIM=0.2887. \textbf{Right:} Fragments of    images. Columns: First - clean fragments; Second -- degraded; Third--   restored by  DAS-2; Fourth -- restored by \textbf{M2}}
\label{mand3_50}
\end{figure}

 Figure  \ref{mand3_50}  displays the restoration results of the ``Mandrill" image, which was  degraded by the application of mask3 (50\% of pixels are missing) and strong additive  Gaussian noise with $\sigma=50$ dB, by  \textbf{M2} and DAS-2, which has produced the highest PSNR from the SET-4 algorithms. Both the PSNR and SSIM values for restoration by   \textbf{M2} are higher than the values   produced by DAS-2.  The  image   restored by DAS-2  looks  over-smoothed compared to restoration by \textbf{M2}. Much of  the fine  texture details, which are lost by the application of DAS-2,  are  retained by \textbf{M2}.

\subsection{A special example}\label{sec:ss44}
\begin{figure}
\centering
\includegraphics[width=3.in]{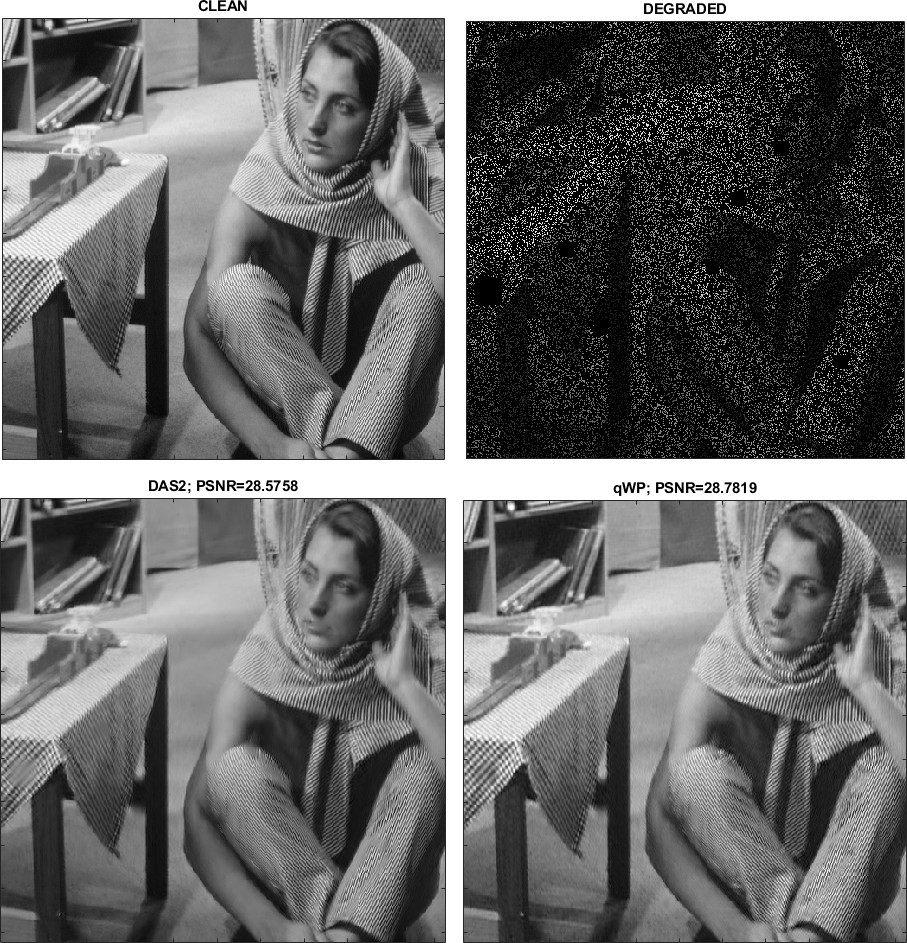}\hfill \includegraphics[width=3.2in]{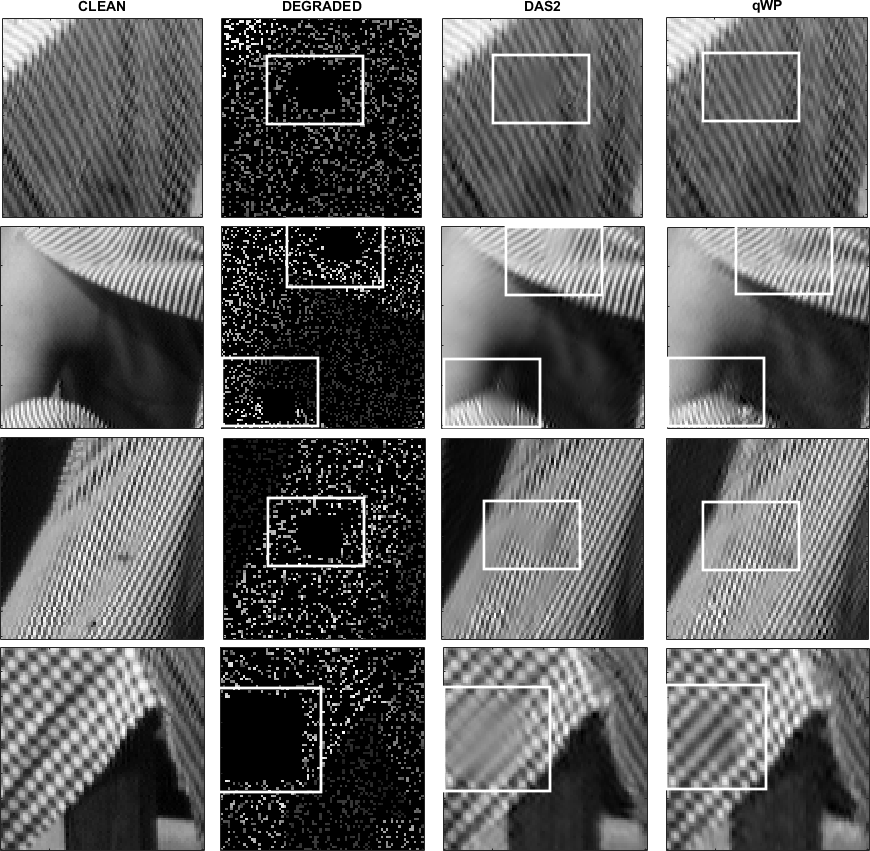}
\caption{Restoration of ``Barbara" image. \textbf{Left:} Top left: clean image. Top right: image degraded by application of modified mask4 (80\% of pixels  and square patches are missing).
Bottom: Restoration by \textbf{M2} (right), PSNR=28.78 dB,  SSIM=0.7312 and by  DAS-2  (left), PSNR=28.58 dB, SSIM=0.7471. \textbf{Right:}  Image fragments. Columns: First - clean fragments; Second -- degraded; Third--   restored by  DAS-2; Fourth -- restored by \textbf{M2}. Bottom row -- gap of size $32\times32$, in the rest of rows -- gaps of size  $16\times16$}
\label{barM8_4}
\end{figure}

All the above experiments were conducted with the inpainting masks shown in Fig. \ref{clima} where missing pixels are either arranged as texts and curves or are randomly distributed. However, our qWP-based algorithms are able to inpaint images where large patches are missing. In the following experiment with ``Barbara'" image, mask4 (80\% of pixels missing) was modified by addition of several $16\times16$ missing squares and one $32\times32$ missing square. Noise was not added. The  \textbf{M2} performance was compared with the performance  of SET-4 algorithms, from which DAS-2 appeared the best. Figure \ref{barM8_4} illustrates results of this experiment. We can see in Fig.  \ref{barM8_4} that, while DAS-2 simply fills the gaps,  \textbf{M2} restores missing complicated structures.

\section{Discussion}\label{sec:s5} The paper presents two  methods   \textbf{M1} and  \textbf{M2} for image impainitng. These methods are based on  the directional quasi-analytic   WPs (qWPs) originated from polynomial  splines of arbitrary order that are designed in \cite{azn_pswq}.  The computational scheme in \textbf{M1} is a slight modification of the  inpainting \textbf{AlgorithmI} introduced in \cite{braver1}.  \textbf{M2}   couples \textbf{M1} with  the Split Bregman Iterations (SBI) scheme. In essence, it is the  SBI algorithm  supplied with  decreasing localized thresholds that are  determined by BSA.

The successful application  of  \textbf{M1} and  \textbf{M2} to image inpainting stems from the exclusive properties of  qWPs   such as:
\begin{description}
\item[-] The qWP transforms  provide a variety of 2D waveforms oriented in multiple directions. For example, fourth-level qWPs are oriented in 62 different directions.
\item[-] The waveforms  are close to directional cosines with a variety of frequencies  that are modulated by spatially  localized low-frequency 2D signals.
\item[-] The waveforms  may have any number of local vanishing moments.
\item[-] The DFT spectra of the waveforms produce a refined tiling of the frequency  domain.
\item[-] The  transforms have a number of free parameters that makes it possible  a flexible adaptation to objects under processing.
\item[-] Fast implementation of the transforms by using FFT that enables us to use the transforms with increased redundancy.
\end{description}
In multiple experiments, the performances of \textbf{M1} and \textbf{M2} are compared with the performances of the state-of-the-art inpainting  algorithms that use the SET-4 filter banks. The inpainting results for 4 images by the SET-4 algorithms, which are degraded by the application of 4 masks (Fig. \ref{clima}) and additive Gaussian  noise of various intensities (noise STD $\sigma=0, 5, 10, 30, 50$ dB), are presented in \cite{che_zhuang}. We added two images, which were not used in \cite{che_zhuang}. For each triple \emph{image-mask-$\sigma$}, the inpainting results by \textbf{M1} and \textbf{M2} vs. the SET-4 algorithms are compared through PSNR, SSIM values and visual inspection.   In all the experiments, our methods succeeded in inpainting  even in severely degraded images and in suppressing noise. In most cases, fine structures were recovered. On images with complicated structures such as  ``Barbara", ``Fingerprint" and ``Mandrill",   \textbf{M1} and especially  \textbf{M2}   significantly outperform the SET-4 algorithms in all three criteria - PSNR, SSIM and visual perception. As for the ``Lena",  ``Boat"  and ``Hill" images, while in most experiments  the SET-4 algorithms produce  higher PSNR values compared to  \textbf{M1} and \textbf{M2}, the SSIM values are higher for \textbf{M1} and especially  for \textbf{M2}. This fact indicates that  our \textbf{M1} and \textbf{M2} are superior over the SET-4 algorithms for image structure recovery. Visual observations  that confirm this claim are illustrated in Figs.  \ref{barb2_50_4} -- \ref{mand3_50}.  Note that  \textbf{M2} succeeded in the restoration of the image fine  structures even in the situations where large patches in the image such as  $16\times16$  and $32\times32$  squares were missing. It is illustrated in Fig. \ref{barM8_4}.

\begin{SCfigure}
\centering
\caption{Restoration of the ``Mandrill" image.  Top left: clean image. Top right: image degraded by the application of mask4 with additive  Gaussian noise with $\sigma=50$ dB.
Bottom right: \textbf{M2} restoration, \underline{PSNR=18.93 dB, SSIM=0.2343}. Bottom left: DAS-2 restoration, \underline{PSNR=19.81 dB, SSIM=0.1414 } }
\includegraphics[width=3.5in]{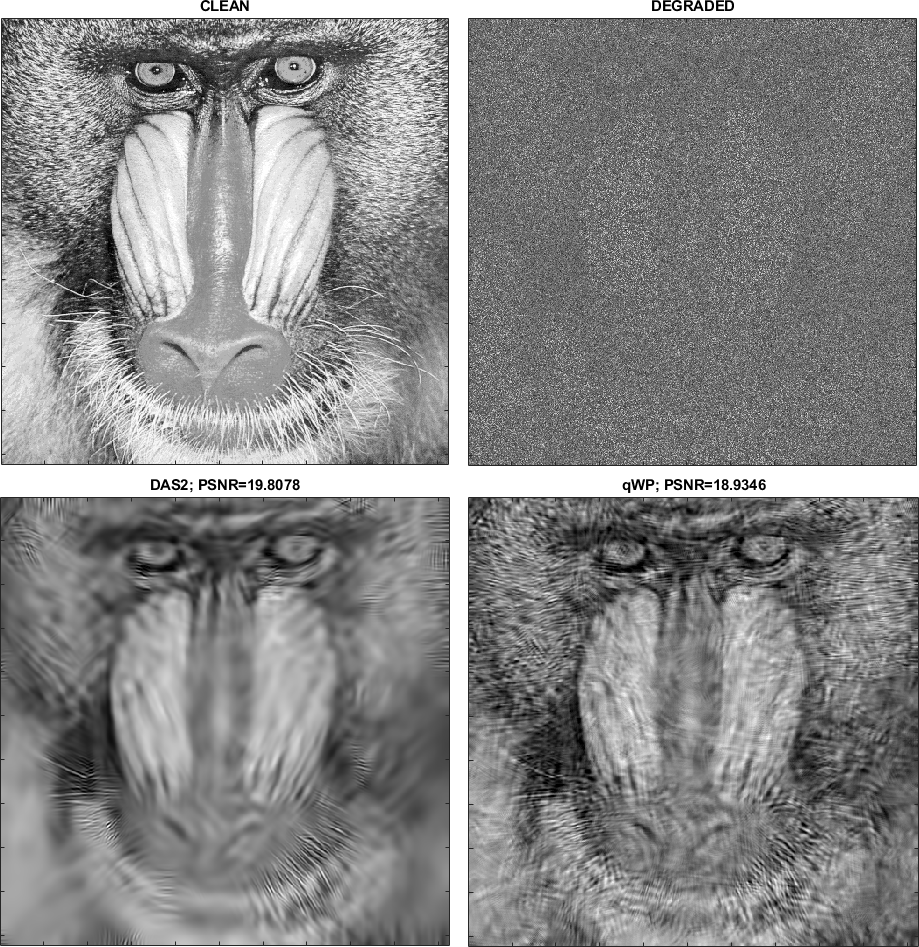}
\label{mand4_50}
\end{SCfigure}

Generally, we claim that our \textbf{M1} and especially  \textbf{M2}, which are based on directional quasi-analytic   wavelet packets originated from polynomial  splines, are powerful tools for images' restoration that are severely degraded by loss of the majority of the pixels and have strong additive noise. These methods succeed in capturing  fine details in the images in cases  where other state-of-the-art algorithms fail. This fact is  illustrated in Fig. \ref{mand4_50}, which displays the restoration results  of the `Mandrill" image from the input where 80\% of pixels are missing and additive Gaussian noise with $\sigma=50$ dB is present.  The output from  DAS-2 algorithm  (which was the best for this image in SET-4) has PSNR=19.81 dB compared to 18.93  dB produced by \textbf{M2}. On the other hand,  the SSIM from the \textbf{M2}  restoration is 0.2343 compared to 0.1414 produced by DAS-2.
 We can see that many texture patterns that were blurred by DAS-2, are restored by  \textbf{M2}. This figure   is a good illustration to the fact that SSIM  is a much more informative characteristics than PSNR.

Summarizing, by having such a versatile and
flexible tool at hand,
we are in a position to address
several data processing problems such as image denoising, deblurring,  superresolution, segmentation and classification, target detection (here the directionality is of utmost importance). The  3D directional wavelet packets, whose design is  underway, may be beneficial for seismic and  hyper-spectral processing.

\paragraph{Acknowledgment}
This research was partially supported by the Israel Science Foundation (ISF, 1556/17),
Supported by Len Blavatnik and the Blavatnik Family Foundation,
Israel Ministry of Science Technology and Space 3-16414, 3-14481 and by Academy of Finland (grant 311514).


\bibliographystyle{plain}
\bibliography{BookBib_TBSIA3}
\end{document}